\documentclass[aps, twocolumn, superscriptaddress]{revtex4-2}

\usepackage{amsmath,amssymb,bm}
\usepackage{physics}
\usepackage{graphicx}
\usepackage{orcidlink}
\newcommand{\safeincludegraphics}[2][]{%
  \IfFileExists{#2}{\includegraphics[#1]{#2}}{%
    \fbox{\parbox{0.88\columnwidth}{Missing figure file: \texttt{\detokenize{#2}}}}%
  }%
}

\usepackage{booktabs}

\begin{document}

\title{Purcell-Engineered Hybrid Coupler for Leakage-Suppressed Robust CZ Gates}
\author{Hui Wang}
\email{wanghuiphy@126.com}
\affiliation{Zhejiang Qizhen Quantum Technology Co, Hangzhou, China}

\author{Feng Bao}
\affiliation{Institute of Quantum Sensing, Zhejiang University, Hangzhou 310027, China}

\author{Yan-Jun Zhao}
\affiliation{Key Laboratory of Opto-electronic Technology, Ministry of Education, Beijing University of
Technology, Beijing, 100124, China}

\author{Xun-Wei Xu}
\affiliation{Key Laboratory of Low-Dimensional Quantum Structures and Quantum Control of Ministry of Education,
 Key Laboratory for Matter Microstructure and Function of Hunan Province,
  Department of Physics and Synergetic Innovation Center for Quantum Effects and Applications,
   Hunan Normal University, Changsha 410081, China}

\begin{abstract}
We propose a Purcell-engineered notch-filter hybrid coupler for
superconducting controlled-$Z$ (CZ) gates that combines coherent
interaction engineering with leakage-selective dissipation. The
architecture integrates a nonlinear transmon coupler with a coupled
Purcell-filter and notch-resonator subsystem, providing additional
control over both the coherent interaction pathways and the engineered
dissipative environment.
The filter branch reshapes the effective interaction pathways,
while the notch resonator further tailors the frequency response
of the coupled filter network and preserves strong
leakage-selective dissipation.
  Using dressed-eigenstate
analysis together with Lindblad master-equation simulations, we show
that the proposed architecture substantially reduces leakage and
retains high-fidelity, low-leakage operation compared with an
optimized single-transmon coupler while remaining robust over a broad
range of coherence assumptions and device parameters.  The gate analysis reconstructs the projected computational quantum channel and
therefore retains the phase information required to distinguish a CZ
operation from identity.  The nominal operating point occurs at
$t_{\rm CZ}=327.46~\mathrm{ns}$ with
$F_{\rm avg}^{\rm CZ}=99.714\%$,
$L_{\rm avg}=8.78\times10^{-4}$, and
$L_{\rm max}=2.04\times10^{-3}$.  A minimax timing-robust operating
point at $327.20~\mathrm{ns}$ maintains
$F_{\rm avg}^{\rm CZ}\geq99.590\%$ over a
$\pm0.50~\mathrm{ns}$ timing window.
These results demonstrate that engineered dissipation complements
conventional coherent interaction engineering and provides an additional
design degree of freedom for realizing robust, high-fidelity
superconducting CZ gates.

\end{abstract}

\date{\today}

\maketitle

\section{Introduction}

Superconducting quantum processors have emerged as one of the leading
platforms for scalable quantum computing owing to their compatibility
with modern microfabrication technologies, fast gate operations, and
continuous improvements in coherence and control
~\cite{Blais2004,Koch2007,Krantz2019,Arute2019,
Acharya2023SurfaceCode,Acharya2025BelowThreshold}.
Steady advances in materials, device design, and quantum control have
enabled single- and two-qubit gate fidelities approaching the thresholds
required for fault-tolerant quantum computation
~\cite{Barends2014,Yan2018,Sung2021}.
Among the available entangling operations, the controlled-phase (CZ)
gate has become a fundamental primitive in superconducting quantum
processors because of its compatibility with fixed-frequency and
tunable-qubit architectures. High-performance CZ gates are commonly
implemented using auxiliary coupling elements, including tunable
transmon couplers, flux-biased couplers, and multi-mode coupling
networks
~\cite{Chen2014,Sung2021,Xu2020ScalableGate,Goto2022DoubleTransmon,Li2024DoubleTransmonCZ,
Wang2024DoubleResonator,HWang,
Campbell2023ModularCoupler,
Li2025PulseCalibration}.
These couplers are designed to generate sufficiently strong
conditional-phase interactions while suppressing unwanted transverse
coupling, residual static interactions, and frequency-crowding effects.
More recently, multi-channel coupling architectures based on multiple
resonators or transmons have attracted increasing interest because they
provide additional degrees of freedom for improving gate performance and
robustness~\cite{Goto2022DoubleTransmon,Li2024DoubleTransmonCZ,
Wang2024DoubleResonator,HWang}.
As coherent control errors continue to decrease, leakage into
non-computational states has become an increasingly important limitation
to high-fidelity quantum operations
~\cite{Motzoi2009,Gambetta2011}.
Leakage is particularly detrimental to quantum error correction because
it can persist over multiple correction cycles and propagate through
entangling gates, creating correlated error channels that are difficult
to mitigate
~\cite{Yang2024LeakageReduction,Ghosh2013,Ghosh2015,Suchara2015,McEwen2021,Wood2018}.
Existing leakage-mitigation strategies primarily rely on pulse shaping,
optimal control, and coherent interaction engineering
~\cite{Motzoi2009,Gambetta2011}.
By comparison, engineered dissipation remains relatively
unexplored as a means of suppressing gate leakage, despite
its demonstrated success in stabilizing quantum states
\cite{Murch,Shankar,Leghtas}
and in protecting superconducting qubits from radiative
decay through Purcell engineering
\cite{Purcell1946,Reed2010,Jeffrey2014,Bronn2015,Sete2015}.
In particular, Purcell filters are widely used to suppress
spontaneous emission through the readout environment while
maintaining fast qubit measurement
~\cite{Sete2015,Reed2010,Bronn2015,DiCarlo2010,Bronn,Jeffrey2014}.
These advances motivate extending dissipation engineering
from qubit protection to the selective suppression of
leakage-related transitions.

In this work, we propose a hybrid coupler architecture consisting of a
tunable transmon coupler, a Purcell-filter resonator and an auxiliary notch resonator.
The Purcell-filter mode couples preferentially to
leakage-related dressed states associated with the
$|1\rangle\!\rightarrow\!|2\rangle$ transitions, whereas the
auxiliary notch mode reshapes the environmental response to suppress
filter-induced decay within the computational manifold.
The resulting filter--notch subsystem enables simultaneous engineering
of coherent CZ interactions and leakage-selective dissipation.
Using dressed-eigenstate analysis to elucidate the microscopic
 leakage-suppression mechanism together with Lindblad master-equation simulations to evaluate CZ-gate
performance, we compare the proposed hybrid architecture with a
conventional single-transmon coupler.
The results demonstrate reduced leakage, improved minimum
computational-state population retention, and robust performance over a practical
parameter range, indicating that engineered dissipation provides an
additional design resource for superconducting CZ gates.

The remainder of this paper is organized as follows.
Section~II introduces the proposed hybrid coupler architecture and
develops the theoretical framework, including the system Hamiltonian,
coherent interaction engineering, and dissipation engineering.
Section~III presents the quantum simulation framework together with the
microscopic leakage-suppression mechanism, CZ-gate performance, and
robustness analyses.
Section~IV discusses the physical implications, practical
considerations, and limitations of the proposed approach.
Finally, Sec.~V summarizes the main conclusions.

\begin{figure}[t]
\centering
\safeincludegraphics[width=1.0\columnwidth]{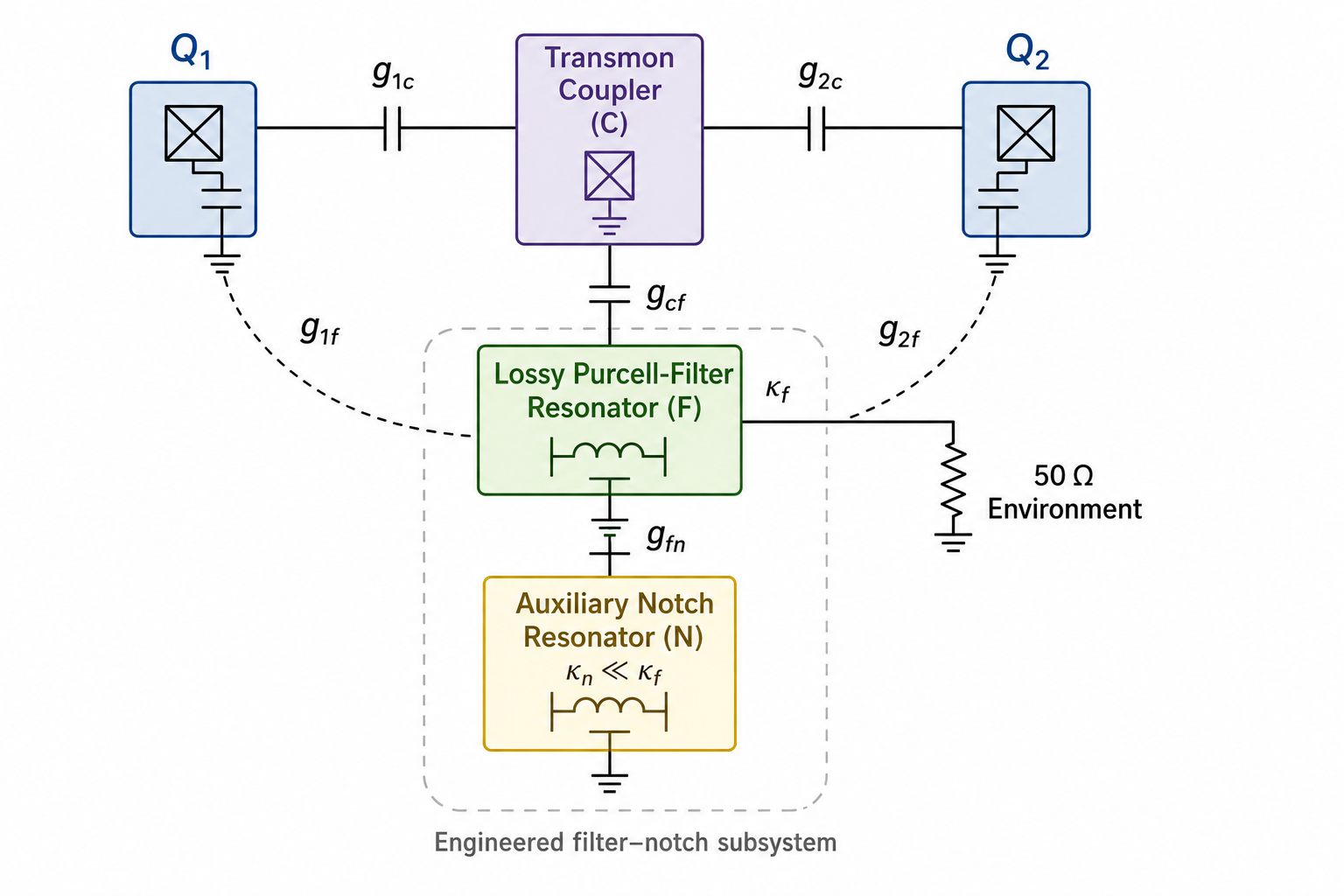}
\caption{\label{fig1}
Schematic of the Purcell-engineered filter--notch
hybrid coupler. The engineered environment comprises two coupled
resonant elements: a lossy Purcell-filter resonator $F$ and a weakly
dissipative auxiliary notch resonator $N$. The filter mode couples to
the dominant $50~\Omega$ dissipative environment with rate $\kappa_f$,
whereas the notch mode couples coherently to the filter through
$g_{fn}$ and has only weak intrinsic loss $\kappa_n$. Together, the two
resonators form a single integrated filter--notch subsystem. Their
interference suppresses the environmental response near the
computational transitions while retaining a stronger response near
selected leakage-related transitions.
}
\end{figure}

\section{Physical Model}

\subsection{Circuit Hamiltonian}

The proposed circuit consists of two frequency-tunable transmon qubits,
a frequency-tunable nonlinear transmon coupler, and an engineered
filter--notch subsystem [Fig.~\ref{fig1}]. The latter comprises two
coupled resonant elements: a lossy Purcell-filter resonator and a
weakly dissipative auxiliary notch resonator. Although represented as
distinct bosonic modes in the Hamiltonian, these elements form a single
integrated filter--notch subsystem.

The annihilation operators of qubits~1 and~2 are denoted by $b_1$ and
$b_2$, respectively. The nonlinear transmon coupler, Purcell-filter
resonator, and auxiliary notch resonator are described by the
annihilation operators $c$, $a$, and $d$. Setting $\hbar=1$, the system Hamiltonian is
$H=H_0+H_{\rm int}$,
where the bare Hamiltonian is
\begin{align}
H_0  ={}&
\sum_{j=1}^{2}
\left[
\omega_j b_j^\dagger b_j
+\frac{\alpha_j}{2} b_j^\dagger b_j^\dagger b_j b_j
\right]
+\omega_c c^\dagger c
+\frac{\alpha_c}{2}c^\dagger c^\dagger c c
\nonumber\\
&+\omega_f a^\dagger a
+\omega_n d^\dagger d .
\label{eq:bare_hamiltonian}
\end{align}
The parameters $\omega_j$ and $\alpha_j$ denote the frequency
and anharmonicity of qubit~$j$, respectively, while $\omega_c$
and $\alpha_c$ denote those of the nonlinear transmon coupler.
The Purcell-filter and notch-resonator frequencies are denoted
by $\omega_f$ and $\omega_n$, respectively.

The interaction Hamiltonian is
\begin{align}
H_{\rm int} ={}&
\sum_{j=1}^{2} g_{jc}\left(b_j^\dagger c+b_j c^\dagger\right)
+g_{12}\left(b_1^\dagger b_2+b_1 b_2^\dagger\right)
\nonumber\\
&+g_{cf}\left(c^\dagger a+c a^\dagger\right)
+\sum_{j=1}^{2}g_{jf}\left(b_j^\dagger a+b_j a^\dagger\right)
\nonumber\\
&+g_{fn}\left(a^\dagger d+a d^\dagger\right).
\label{eq:interaction_hamiltonian}
\end{align}
The interaction term $g_{jc}$ couples qubit $j$ to the nonlinear
transmon coupler~\cite{Blais2004,Yan2018}, while $g_{12}$ denotes the
direct qubit--qubit coupling. The coupling $g_{cf}$ connects the
nonlinear coupler to the Purcell-filter resonator, and the weak
qubit--filter couplings $g_{jf}$ provide an additional exchange pathway
that modifies the effective transverse interaction.
The interaction $g_{fn}$ couples the Purcell-filter resonator to the
auxiliary notch mode, introducing an additional resonant pole into the
frequency response of the coupled filter--notch subsystem.
Because the notch resonator couples only to the filter, it reshapes the
electromagnetic environment without introducing an additional direct
qubit-coupling channel.

\begin{figure}[t]
\centering
\safeincludegraphics[width=0.8\columnwidth]{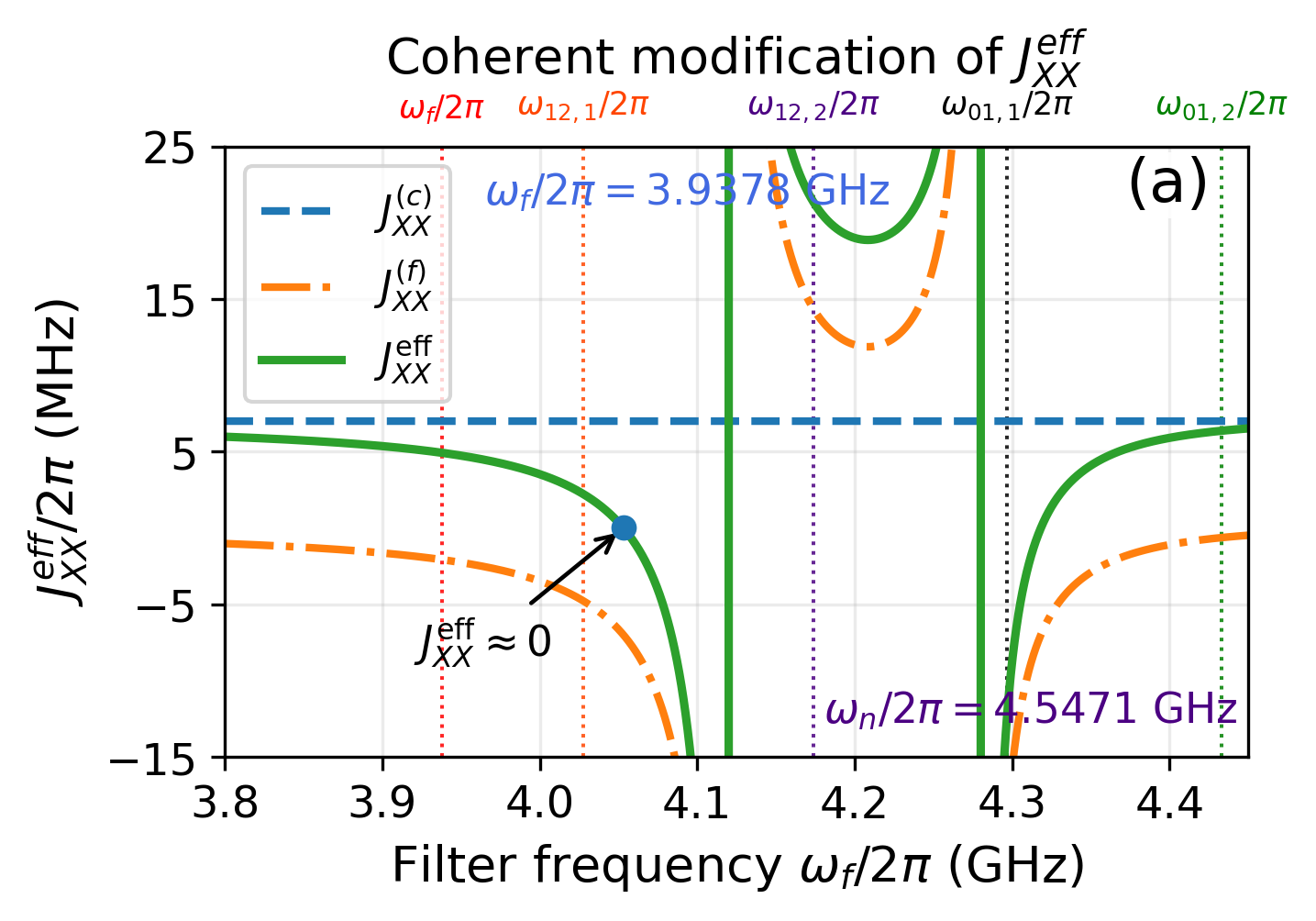}
\safeincludegraphics[width=0.8\columnwidth]{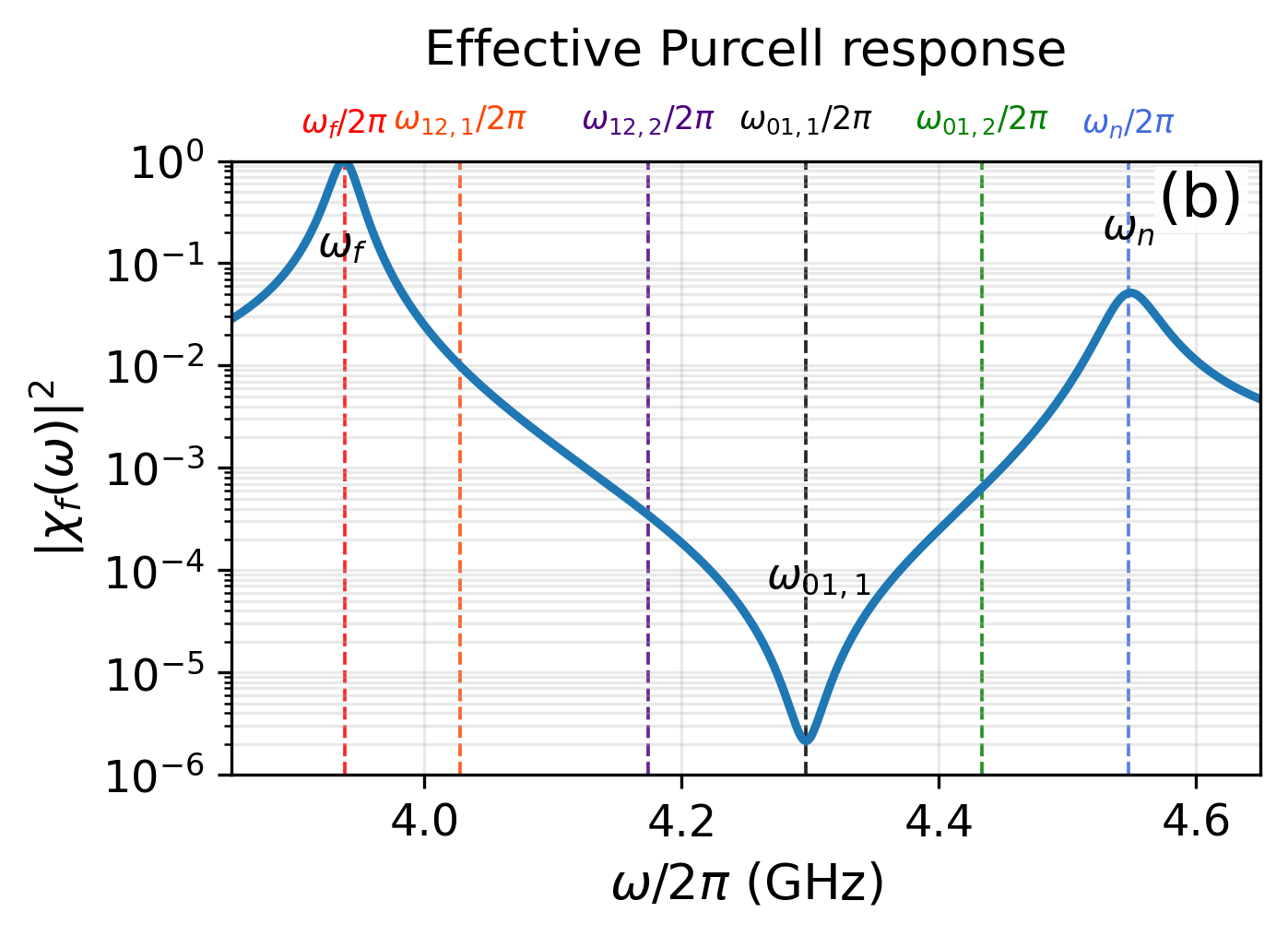}
\caption{\label{fig2}
(a) Representative dependence of the effective transverse interaction
$J_{XX}^{\rm eff}$ on the filter frequency $\omega_f$. The effective
interaction consists of the direct qubit--qubit coupling $g_{12}$, the
virtual exchange mediated by the nonlinear transmon coupler,
$J_{XX}^{(\mathrm{c})}$ [Eq.~(\ref{eq:jxx_coupler})], and the
filter-mediated contribution $J_{XX}^{(\mathrm{f})}$
[Eq.~(\ref{eq:jxx_filter})], giving
$J_{XX}^{\rm eff}\simeq
g_{12}+J_{XX}^{(\mathrm{c})}+J_{XX}^{(\mathrm{f})}$.
The near-zero crossing is included only to illustrate the available
tuning range and does not represent the final CZ operating point.
(b) Qualitative frequency response
$R(\omega)=|\chi_f(\omega)|^2$ of the coupled
filter--notch subsystem [Eq.~(\ref{eq:chi_filter})].
Throughout this figure, $\omega_{01,j}$ and
$\omega_{12,j}$ ($j=1,2$) denote the bare
$0\!\rightarrow\!1$ (computational) and
$1\!\rightarrow\!2$ (leakage) transition
frequencies of qubit $j$.
}
\end{figure}

In a practical coplanar-waveguide implementation, the designed
filter--notch coupling can be realized by positioning a voltage
antinode of the notch resonator adjacent to a voltage antinode of the
Purcell-filter resonator and introducing a controlled coupling
capacitance. Direct notch--qubit coupling can be minimized by spatially
separating the notch resonator from the qubit capacitor pads and, if
necessary, by incorporating ground shielding or via fences. The
resulting hierarchy, $g_{fn}\gg g_{1n},g_{2n}$
justifies treating the notch mode as coupled to the qubits predominantly
through the Purcell-filter mode.
Compared with a conventional transmon coupler, the coupled
filter--notch subsystem introduces additional degrees of freedom for
simultaneously engineering coherent interactions and the dissipative
environment.

\subsection{Coherent Interaction Engineering}

Besides governing SWAP-type interactions, the residual transverse
coupling also influences leakage during CZ-gate operation.
The nonlinear transmon coupler provides the
dominant coherent interaction responsible for the CZ gate, while the
filter branch introduces an additional exchange pathway that reshapes
the residual transverse interaction. To leading order, the effective transverse interaction can
be decomposed into three contributions, $J_{XX}^{\rm eff}
\simeq
g_{12}
+
J_{XX}^{(\rm c)}
+
J_{XX}^{(\rm f)}$.
And $g_{12}$ denotes the direct exchange coupling between the two
qubits,
$J_{XX}^{(\rm c)}$ is the virtual exchange interaction mediated by the
nonlinear transmon coupler, and
$J_{XX}^{(\rm f)}$ is the additional contribution arising from the weak
qubit--filter couplings. To leading order, the virtual exchange mediated by the nonlinear
transmon coupler is
\begin{equation}
J_{XX}^{(\rm c)}
=
\frac{g_{1c}g_{2c}}{2}
\left(
\frac{1}{\Delta_{1c}}
+
\frac{1}{\Delta_{2c}}
\right),
\label{eq:jxx_coupler}
\end{equation}
where
$\Delta_{jc}=\omega_j-\omega_c$.

Treating the Purcell filter as an approximately linear resonator, the
filter-mediated contribution can be written as~\cite{Gong2021,Wang2024DoubleResonator,HWang,Wang2022,Wallraff}

\begin{equation}
J_{XX}^{(\rm f)}
\simeq
\frac{g_{1f}g_{2f}}{2}
\left(
\frac{1}{\Delta_{1f}}
+
\frac{1}{\Delta_{2f}}
\right),
\label{eq:jxx_filter}
\end{equation}
with
$\Delta_{jf}=\omega_j-\omega_f$.
The direct coupling is fixed by the circuit layout, whereas the
transmon-coupler and filter-mediated contributions can be tuned through
their respective detunings.

Figure~\ref{fig2}(a) illustrates this coherent tuning mechanism.
Over the plotted range, $J_{XX}^{(\mathrm{c})}$ is nearly constant,
whereas $J_{XX}^{(\mathrm{f})}$ varies strongly with filter detuning,
allowing $J_{XX}^{\rm eff}$ to be tuned through $\omega_f$.
The near-zero crossing illustrates the available tuning range and is
not the final CZ operating point. The role of the filter is not merely to force $J^{eff}_{XX}$
 exactly to zero, but to provide an additional degree of freedom
  for balancing coherent interactions, leakage suppression, and gate fidelity.
 The analytical expressions above
provide qualitative physical insight, whereas all numerical results
presented in this work are obtained from full Hamiltonian
diagonalization unless stated otherwise.

The gate duration is primarily determined by the dressed-state
interaction strength $J_{ZZ}$, extracted from the dressed eigenenergies
as
\begin{equation}
J_{ZZ}=E_{11}-E_{10}-E_{01}+E_{00}.
\end{equation}
Here $E_{ij}$ are the dressed eigenfrequencies associated with
the computational states. Throughout this work, extracted
interaction strengths are reported in ordinary-frequency units
(MHz), obtained by dividing the corresponding angular-frequency
quantities by $2\pi$.

\subsection{Dissipation Engineering and Purcell Protection}

The coupled filter--notch subsystem also engineers the
frequency-dependent electromagnetic environment experienced by the
qubits. Because the notch resonator couples only to the Purcell-filter
resonator through $g_{fn}$, it reshapes the frequency response of the
coupled filter--notch subsystem without directly interacting with the
qubits. Solving the linear coupled-mode equations in the frequency
domain yields \cite{Sete2015,Bronn,Gardiner1985}
\begin{equation}
\chi_f(\omega)
=
\frac{1}
{
(\omega-\omega_f)+i\kappa_f/2
-\Sigma_n(\omega)
},
\label{eq:chi_filter}
\end{equation}
where
$\Sigma_n(\omega)=g_{fn}^{\,2}/
(\omega-\omega_n+i\kappa_n/2)$
is the self-energy correction induced by the notch resonator.
The corresponding frequency response is characterized by
$R(\omega)=|\chi_f(\omega)|^2$, which provides a qualitative measure
of the coupling between circuit transitions and the engineered
environment. Through interference with the lossy filter pathway, the
notch resonator acts as a frequency-selective protection element. It
suppresses the environmental response near the computational
$0\!\rightarrow\!1$ transitions, thereby reducing filter-induced
Purcell relaxation, while retaining a stronger response near selected
leakage-related transitions.

Figure~\ref{fig2}(b) illustrates a representative frequency response described
by Eq.~(\ref{eq:chi_filter}), evaluated using representative
parameters chosen to highlight the qualitative spectral features of the
coupled filter--notch subsystem.
The peak near $\omega_f$ originates from the Purcell-filter resonance,
whereas the feature near $\omega_n$ is introduced by the notch
resonator. The response exhibits a spectral minimum positioned near a
representative computational transition, illustrating the desired
suppression of the environmental response for computational
excitations. In contrast, representative leakage-related transitions
remain outside this protected spectral region and therefore
experience a stronger effective environmental response in this
qualitative frequency-domain picture~\cite{Jeffrey2014,Sete2015,Bronn}.

\section{Quantum Simulation}

In the optimized operating regime, we mainly  try to suppress leakage and
improve gate robustness for  quantum simulation. The gate performance is evaluated using Lindblad
master-equation simulations~\cite{Lindblad,Gorini},
\begin{equation}
\dot{\rho}
=
-i[H,\rho]
+
\sum_k \mathcal{D}[L_k]\rho,
\label{eq:master_equation}
\end{equation}
where
$\mathcal{D}[L_k]\rho
=
L_k\rho L_k^\dagger
-
\frac{1}{2}
\{L_k^\dagger L_k,\rho\}$.
The collapse operators are
$L_{1,j}=\sqrt{\gamma_{1,j}}\,b_j$,
$L_{\phi,j}=\sqrt{\gamma_{\phi,j}}\,b_j^\dagger b_j$,
$L_{1,c}=\sqrt{\gamma_{1,c}}\,c$,
$L_{\phi,c}=\sqrt{\gamma_{\phi,c}}\,c^\dagger c$,
$L_f=\sqrt{\kappa_f}\,a$, and
$L_n=\sqrt{\kappa_n}\,d$, where $j=1,2$,
$\gamma_{1,j}=1/T_{1,j}$,
$\gamma_{\phi,j}=1/T_{\phi,j}$,
$\gamma_{1,c}=1/T_{1,c}$, and
$\gamma_{\phi,c}=1/T_{\phi,c}$.
 The numerical simulations are implemented using QuTiP~\cite{Johansson1,Johansson2}.

The system is initialized in
the computational basis and evolves under the time-independent
Hamiltonian corresponding to a fixed gate operating point.
The transient biasing process between the idle and gate operating points
is not modeled explicitly.
For each parameter set, the interaction time is optimized by sweeping
the gate duration, and the procedure is repeated for different device
parameters to evaluate the phase-sensitive CZ fidelity, leakage, and robustness.
This approximation enables a direct comparison of different coupler
architectures while isolating the influence of the engineered
filter--notch environment.
Throughout this work, we distinguish the following state
representations.
The bare occupation-number basis of the full five-mode Hilbert space is written as $|q_1,q_2,n_c,n_f,n_n\rangle$
where $q_1$ and $q_2$ denote the occupation numbers of qubit~1 and qubit~2, respectively, and $n_c$, $n_f$, and $n_n$ denote the excitation numbers of the transmon coupler, Purcell filter, and notch resonator.

\begin{table}[t]
\caption{\label{tab:device_parameters}
Optimized hybrid-coupler parameters used for the local-refinement simulations.}
\begin{ruledtabular}
\begin{tabular}{lc}
Parameter & Value \\
\hline
$\omega_1/2\pi$ & $4.2966~{\rm GHz}$ \\
$\omega_2/2\pi$ & $4.4331~{\rm GHz}$ \\
$\omega_c/2\pi$ & $5.6750~{\rm GHz}$ \\
$\omega_f/2\pi$ & $3.9378~{\rm GHz}$ \\
$\omega_n/2\pi$ & $4.5471~{\rm GHz}$ \\
$\alpha_1/2\pi$ & $-268.8~{\rm MHz}$ \\
$\alpha_2/2\pi$ & $-259.4~{\rm MHz}$ \\
$\alpha_c/2\pi$ & $-285.0~{\rm MHz}$ \\
$g_{1c}/2\pi$ & $29.25~{\rm MHz}$ \\
$g_{2c}/2\pi$ & $31.00~{\rm MHz}$ \\
$g_{12}/2\pi$ & $9.00~{\rm MHz}$ \\
$g_{cf}/2\pi$ & $92.50~{\rm MHz}$ \\
$g_{1f}/2\pi$ & $5.80~{\rm MHz}$ \\
$g_{2f}/2\pi$ & $6.70~{\rm MHz}$ \\
$g_{fn}/2\pi$ & $95.09~{\rm MHz}$ \\
\end{tabular}
\end{ruledtabular}
\end{table}

\begin{table}[t]
\caption{\label{tab:simulation_parameters}
Coherence and loss parameters used in the local-refinement master-equation simulations.}
\begin{ruledtabular}
\begin{tabular}{lc}
Parameter & Value \\
\hline
$T_{1,q}$ & $300~\mu{\rm s}$ \\
$T_{\phi,q}$ & $300~\mu{\rm s}$ \\
$T_{1,c}$ & $100~\mu{\rm s}$ \\
$T_{\phi,c}$ & $80~\mu{\rm s}$ \\
$\kappa_f/2\pi$ & $2.0~{\rm MHz}$ \\
$\kappa_n/2\pi$ & $0.02~{\rm MHz}$ \\
\end{tabular}
\end{ruledtabular}
\end{table}

\begin{table}[t]
\caption{\label{tab:coherence_sets}
Coherence sets used in the sensitivity study.
Set 6 represents an approximate near-lossless reference limit
used to separate coherent leakage from ordinary relaxation and
dephasing effects. All coherence times are in $\mu{\rm s}$.}
\begin{ruledtabular}
\begin{tabular}{ccccc}
Set & $T_{1,q}$ & $T_{\phi,q}$ & $T_{1,c}$ & $T_{\phi,c}$ \\
\hline
1 & 100 & 60 & 50 & 20 \\
2 & 150 & 100 & 75 & 40 \\
3 & 200 & 150 & 100 & 60 \\
4 & 300 & 200 & 100 & 80 \\
5 & 500 & 500 & 200 & 150 \\
6 & 1000 & 1000 & 1000 & 1000 \\
\end{tabular}
\end{ruledtabular}
\end{table}

\textbf{a. Computational basis.}
The computational subspace consists of $\{|00\rangle,|01\rangle,|10\rangle,|11\rangle\}$
with the coupler, Purcell filter, and notch resonator remaining in their ground states. Thus, $|q_1 q_2\rangle \equiv |q_1,q_2,0,0,0\rangle $, where $q_1,q_2\in\{0,1\}$.

\textbf{b. Extended multimode basis.}
The tuple  $(q_1,q_2,n_c,n_f,n_n)$
labels the bare occupation-number basis of the full Hilbert space,
whereas the physical eigenstates are the corresponding dressed
eigenstates of the interacting Hamiltonian.

The optimized coherent parameters defining the reference
operating point are summarized in Table~\ref{tab:device_parameters}, while the
reference coherence and loss parameters are listed in
Table~\ref{tab:simulation_parameters}. The additional coherence sets used in the
sensitivity study are summarized in Table~\ref{tab:coherence_sets}.
Unless otherwise stated, the simulations employ the reference coherence
set
($T_{1,q}=T_{\phi,q}=300~\mu{\rm s}$,
$T_{1,c}=100~\mu{\rm s}$,
$T_{\phi,c}=80~\mu{\rm s}$),
reducing the influence of ordinary relaxation and dephasing and
thereby emphasizing the contribution of the engineered filter--notch
architecture. Additional coherence sets in Table~\ref{tab:coherence_sets} are used to verify that the
observed improvements are robust under experimentally relevant
conditions. The following sections first examine the microscopic
leakage-suppression mechanism through the engineered
decay rates of dressed eigenstates and then evaluate the
resulting CZ-gate performance within the same
master-equation framework.

\subsection{Leakage Analysis}

\subsubsection{Mechanism of Leakage Suppression}

Because coherent interactions hybridize the bare basis states, the
leakage-suppression mechanism is more directly characterized by the
decay rates of the dressed eigenstates than by the bare-frequency
response alone. In the hybrid architecture, leakage-like dressed
states couple more strongly to the engineered dissipative environment,
whereas computational-like dressed states remain comparatively
protected. This selective damping shortens the lifetime of residual
leakage without requiring coherent revival into the computational
subspace.

Figure~\ref{fig2}(b) provides a qualitative frequency-domain picture in which
computational transitions are weakly coupled to the environment,
whereas leakage-related transitions experience stronger effective
dissipation~\cite{Yang2024LeakageReduction,Bronn2015,Purcell1946,Blais},
\begin{equation}
\kappa_{\rm env}(\omega_{01})
\ll
\kappa_{\rm env}(\omega_{12}),
\label{eq:env_target}
\end{equation}
where $\kappa_{\rm env}(\omega)$ denotes the effective
frequency-dependent decay rate produced by the filter--notch
subsystem. Here, $\omega_{01}$ and $\omega_{12}$ denote representative
computational and leakage-related transition frequencies,
respectively.  Equation~(\ref{eq:env_target}) provides a qualitative
frequency-domain design criterion. More generally, the use of
controlled dissipation as a resource is closely related to
reservoir-engineering approaches for stabilizing desired quantum
states \cite{Murch,Shankar,Leghtas}.

We investigate the leakage-suppression mechanism by analyzing the
engineered decay rates of dressed eigenstates.
 If $|\psi_n\rangle$ and $|\psi_m\rangle$ are dressed eigenstates,
the transition rate from $|\psi_n\rangle$ to $|\psi_m\rangle$
induced by the engineered filter and notch dissipation channels is
\begin{equation}
\Gamma_{n\rightarrow m}^{(\mathrm{eng})}
=
\left|
\left\langle\psi_m\left|\sqrt{\kappa_f}a\right|\psi_n\right\rangle
\right|^2
+
\left|
\left\langle\psi_m\left|\sqrt{\kappa_n}d\right|\psi_n\right\rangle
\right|^2 ,
\label{eq:eng_decay_rate}
\end{equation}
where $\kappa_f$ and $\kappa_n$ are the decay rates of the filter
and notch modes, respectively. The total engineered decay rate of
$|\psi_n\rangle$ is obtained by summing over all energetically
allowed downward transitions,
\begin{equation}
\Gamma_n^{(\mathrm{eng})}
=
\sum_{E_m<E_n}
\Gamma_{n\rightarrow m}^{(\mathrm{eng})}.
\label{eq:eng_decay_rate_sum}
\end{equation}

To quantify the selectivity of the engineered dissipation, we define
the mean decay rates of the computational and leakage-related
dressed-state manifolds as $ \Gamma_{\rm comp}
=
\frac{1}{N_{\rm comp}}
\sum_{i\in\mathcal{C}}
\Gamma_i^{(\mathrm{eng})}$  and $\Gamma_{\rm leak}
=
\frac{1}{N_{\rm leak}}
\sum_{j\in\mathcal{L}}
\Gamma_j^{(\mathrm{eng})}$, respectively,
where $\mathcal{C}$ and $\mathcal{L}$ denote the computational and
leakage-related dressed-state manifolds, respectively, and
$N_{\rm comp}$ and $N_{\rm leak}$ are the corresponding numbers of
states.

\begin{figure*}[t]
\centering
\safeincludegraphics[width=0.73\textwidth]{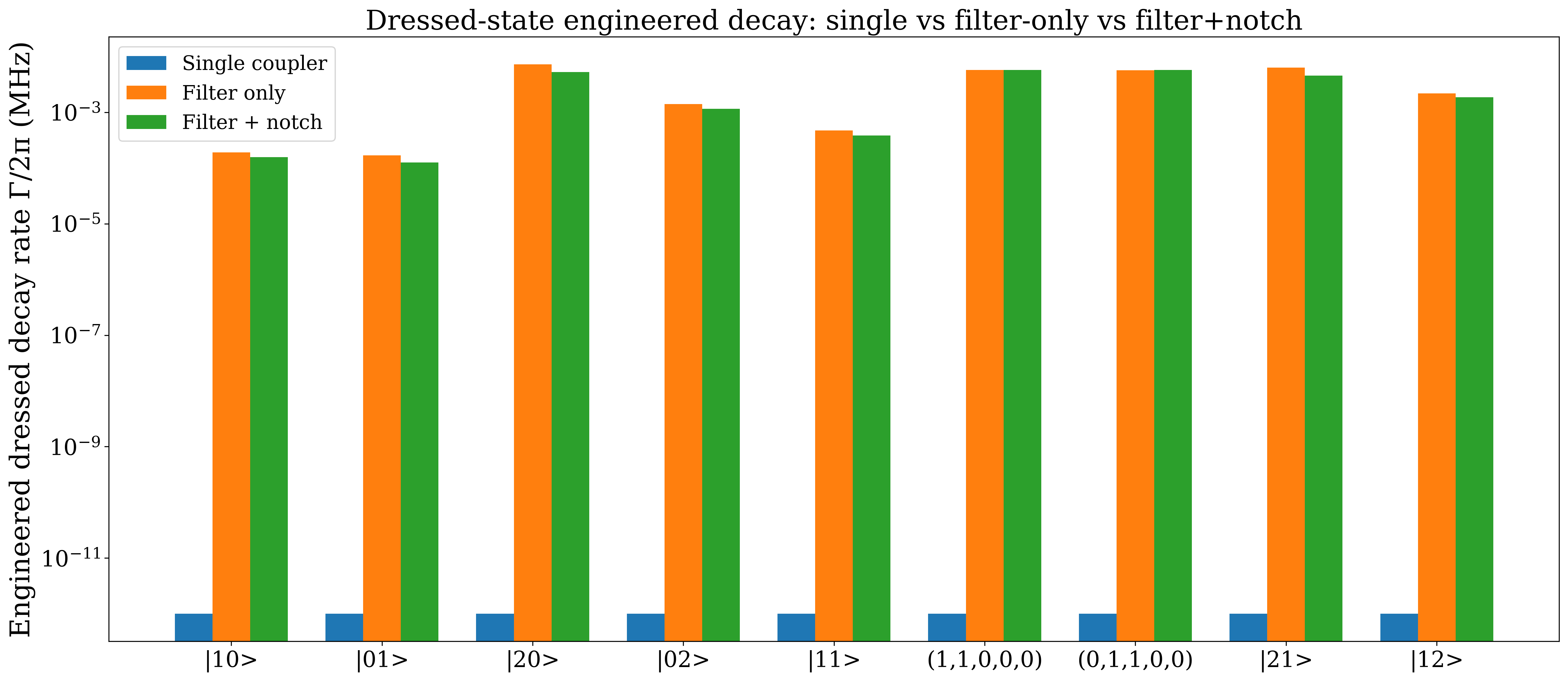}
\caption{\label{fig:dressed_decay_rates}
Engineered  decay rates for representative computational and leakage-related eigenstates. Results are shown for the single-coupler, filter-only ($g_{fn}=0$), and full filter--notch architectures.
Leakage-dominated dressed eigenstates, identified by their largest overlap with the corresponding bare leakage states
$|20\rangle$, $|02\rangle$, $|21\rangle$, and $|12\rangle$, exhibit decay rates more than an order of magnitude larger than those of the computational manifold ($|10\rangle$, $|01\rangle$, and $|11\rangle$), demonstrating strong leakage-selective dissipation.
Tuple labels follow the notation defined in Sec.~II.A. For example, $(1,1,0,0,0)$ corresponds to the bare configuration associated with $|11\rangle$, while $(0,1,1,0,0)$ denotes the bare configuration with one excitation in qubit~2 and one in the coupler mode.
}
\end{figure*}

Figure~\ref{fig:dressed_decay_rates} shows the engineered decay rates
for the single-coupler, filter-only, and full filter--notch
architectures. The filter-only architecture is obtained by setting
$g_{fn}=0$ while leaving the other interactions unchanged.
For the filter-only architecture,
$\Gamma_{\rm comp}=2.78\times10^{-4}$ MHz and
$\Gamma_{\rm leak}=4.11\times10^{-3}$ MHz, whereas the full
filter--notch architecture gives
$\Gamma_{\rm comp}=2.22\times10^{-4}$ MHz and
$\Gamma_{\rm leak}=3.40\times10^{-3}$ MHz.
Thus, $\Gamma_{\rm leak}/\Gamma_{\rm comp}\approx15$ in both
hybrid architectures, demonstrating that leakage-related dressed
states experience approximately an order of magnitude stronger
engineered decay than the computational manifold. This separation
provides the microscopic basis for the reduced bare-projector
leakage discussed below.

\subsubsection{Random-State Bare-Projector Leakage}

The dressed-state analysis in Sec.~III~A explains the microscopic
origin of leakage suppression. To evaluate the resulting CZ-gate performance, however, leakage is
quantified by the population outside the bare computational subspace.
To characterize leakage beyond the four computational basis states, we
consider an ensemble of random initial states within the computational
subspace, $|\psi_0\rangle=c_{00}|00\rangle+c_{01}|01\rangle+c_{10}|10\rangle+c_{11}|11\rangle$,
where the complex coefficients satisfy the normalization condition $\sum_{i,j}|c_{ij}|^2=1$.
Each state is evolved under the Lindblad master equation to obtain the
final density matrix $\rho_f$.
Leakage is quantified by the projector onto the bare computational
subspace embedded in the full Hilbert space,
\begin{equation}
P_{\mathrm{bare}} =
\sum_{q_1,q_2=0,1}
|q_1,q_2,0_c,0_f,0_n\rangle
\langle q_1,q_2,0_c,0_f,0_n|,
\end{equation}

The corresponding bare-projector leakage is
\begin{equation}
L_{\mathrm{bare}}
=
1-\operatorname{Tr}\!\left(P_{\mathrm{bare}}\rho_f\right).
\end{equation}
It is important to distinguish suppression of residual leakage from
recovery of the intended logical state. Population remaining in a
noncomputational state, such as $|02\rangle$, constitutes a leakage
error that may persist and affect subsequent gate or error-correction
cycles. Coupling a leakage-like dressed state to the lossy filter can
shorten its lifetime by inducing transitions to lower-energy dressed
states. Such relaxation removes population from the leakage manifold
but does not generally restore the intended CZ output. For example, a
single-excitation decay of a predominantly $|02\rangle$-like state may
produce a lower-energy state with predominantly $|01\rangle$-like
character, thereby converting a leakage error into a decay-induced
computational error rather than correcting it. The phase-sensitive CZ fidelity evaluated below accounts for both
leakage and decay-induced computational errors, whereas
$L_{\mathrm{bare}}$ measures only the population remaining outside the
specified computational subspace at the end of the gate.

\begin{figure}[t]
\centering
\safeincludegraphics[width=0.75\columnwidth]{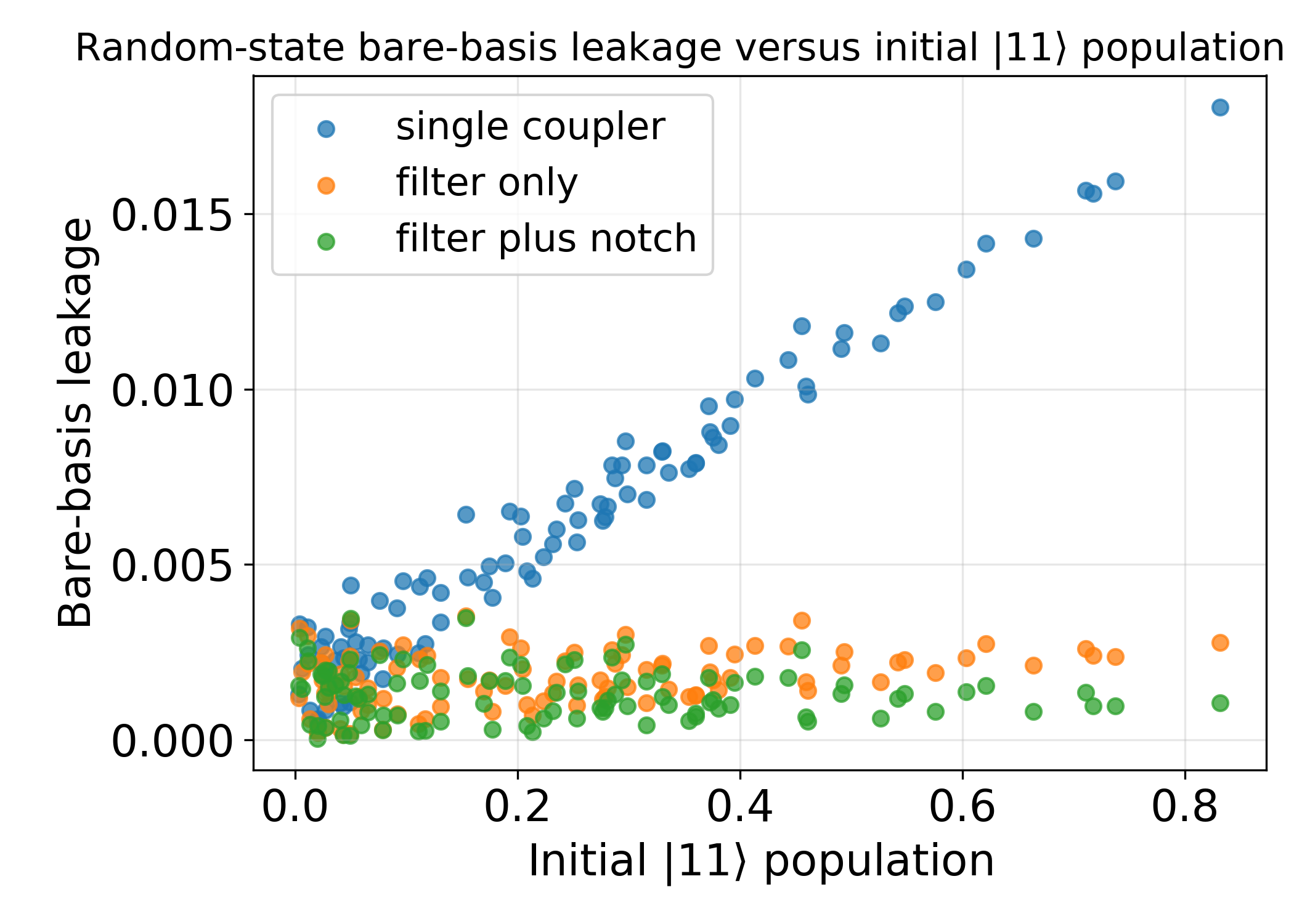}
\caption{
Random-state bare-projector leakage as a function of the initial $|11\rangle$ population.
 Each point corresponds to a different randomly generated computational state
$|\psi_0\rangle=c_{00}|00\rangle+c_{01}|01\rangle+c_{10}|10\rangle+c_{11}|11\rangle$,
with all four complex amplitudes sampled simultaneously and normalized. Thus, the figure represents a statistical correlation between leakage and the initial $|11\rangle$ population rather than a scan performed at fixed $P_{11}$.
}
\label{fig:leakage_P11_correlation_scatter}
\end{figure}

For each realization, the initial two-excitation weight
$P_{11}=|c_{11}|^2$ is also recorded, providing a measure of the
population in the leakage-prone manifold. The dependence of
$L_{\mathrm{bare}}$ on $P_{11}$ therefore probes leakage associated
with transitions such as
$|11\rangle\leftrightarrow|20\rangle$ and
$|11\rangle\leftrightarrow|02\rangle$.
Figure~\ref{fig:leakage_P11_correlation_scatter} shows the correlation
between $L_{\mathrm{bare}}$ and $P_{11}$. As shown in Fig.~\ref{fig:leakage_P11_correlation_scatter}, the
single-coupler architecture exhibits a strong increase of leakage with
$P_{11}$, with a fitted slope of $1.97\times10^{-2}$ and a maximum
leakage of $1.8\times10^{-2}$. This identifies the two-excitation
manifold as the dominant leakage pathway. The filter-only and
filter--notch architectures keep the leakage substantially lower and
strongly reduce its dependence on the initial $|11\rangle$ population.

\begin{figure}[t]
\centering
\safeincludegraphics[width=0.8\columnwidth]{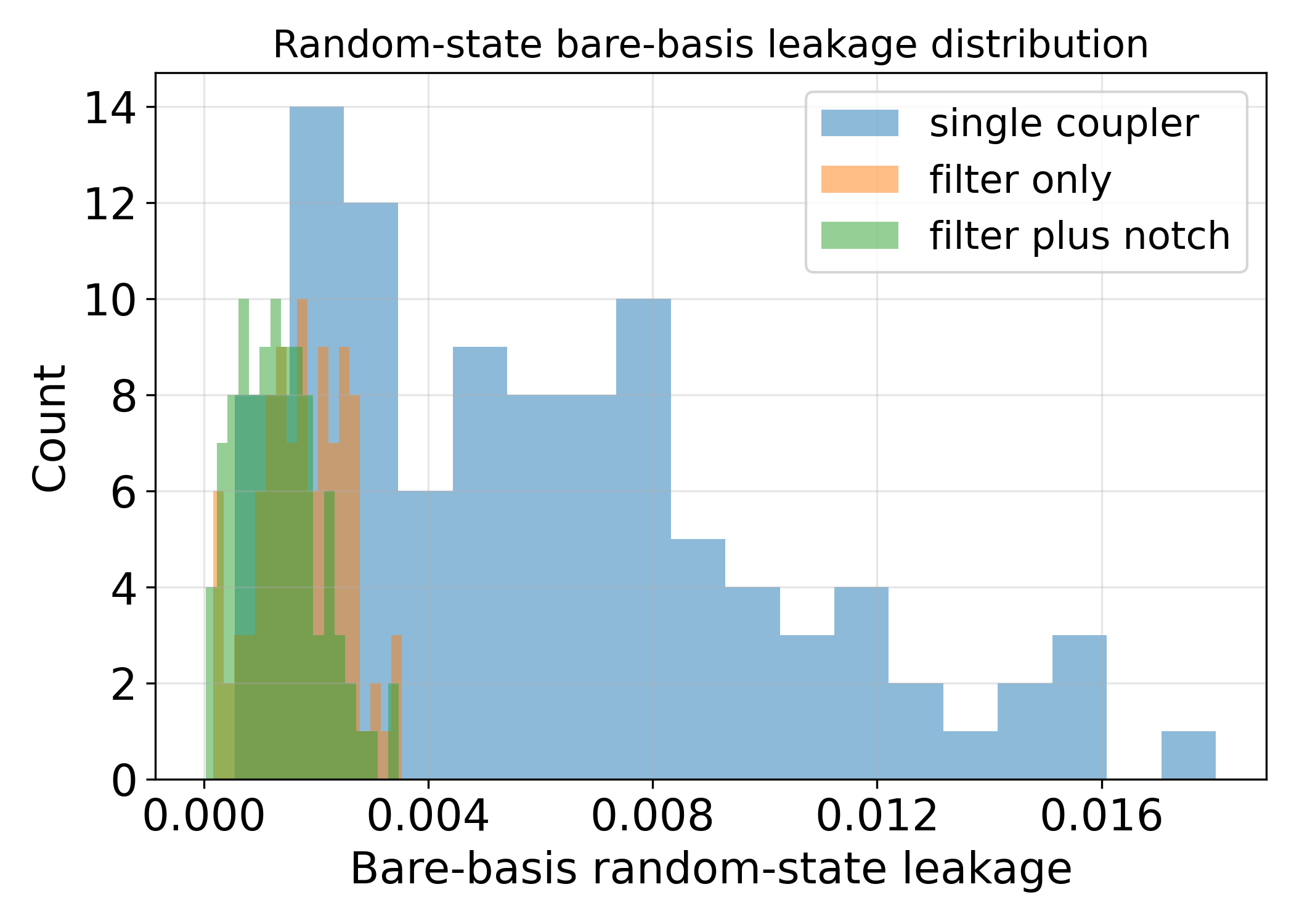}
\caption{
Distribution of random-state bare-projector leakage for the optimized
single-coupler, filter-only, and filter+notch architectures. The hybrid
architectures shift the leakage distribution toward lower values compared with
the single-coupler reference, with the filter+notch design giving the smallest
mean leakage.
}
\label{fig:leakage_distribution_bare_histogram}
\end{figure}

\begin{table}[t]
\centering
\caption{
Random-state bare-projector leakage statistics for the three circuit
architectures. The final column gives the linear slope obtained from fitting
$L_{\rm bare}$ as a function of the initial $\vert 11\rangle$ population.
}
\label{tab:random_state_leakage_bare}
\begin{tabular}{lccc}
\hline\hline
Architecture & Mean leakage & Maximum leakage & Slope vs. $P_{11}$ \\
\hline
Single coupler & $6.15\times10^{-3}$ & $1.80\times10^{-2}$ & $1.97\times10^{-2}$ \\
Filter only & $1.76\times10^{-3}$ & $3.53\times10^{-3}$ & $1.42\times10^{-3}$ \\
Filter+notch & $1.31\times10^{-3}$ & $3.48\times10^{-3}$ & $-4.04\times10^{-4}$ \\
\hline\hline
\end{tabular}
\end{table}

Figure~\ref{fig:leakage_distribution_bare_histogram} illustrates the
leakage distributions, while Table~\ref{tab:random_state_leakage_bare} summarizes the corresponding
statistical measures. Compared with the single-coupler architecture,
the filter-only design reduces the mean bare-projector leakage from
$6.15\times10^{-3}$ to $1.76\times10^{-3}$, while the full
filter--notch architecture further reduces it to
$1.31\times10^{-3}$ and nearly eliminates its dependence on $P_{11}$.
Together with the dressed-state decay analysis in Sec.~III.A, these
results support the interpretation of selective damping of
leakage-like dressed states.

\subsection{CZ-Gate Performance}

Having established the microscopic leakage-suppression mechanism,
we now evaluate the resulting CZ-gate performance.
Unless otherwise stated, leakage denotes the final population outside
the bare computational subspace.
The system evolves under the time-independent gate Hamiltonian for a
duration $t_{\rm CZ}$ according to the Lindblad master equation
[Eq.~(\ref{eq:master_equation})].
After the evolution, independent local-$Z$ phase corrections are
applied numerically before evaluating the gate fidelity~\cite{Sung2021,McKay}.
Throughout this work, the qubits
and coupler are held at fixed operating frequencies during the CZ gate,
such that the gate dynamics are governed by a time-independent
Hamiltonian.

\subsubsection{Local Optimization of the Hybrid Coupler}

The optimization seeks to maximize the conditional interaction
$|J_{ZZ}|$ while suppressing the residual transverse coupling
$J_{XX}^{\rm eff}$ and leakage. The corresponding CZ-gate duration is
initially estimated as
$t_{\rm CZ}\approx1/(2|J_{ZZ}|)$.
Starting from the globally optimized operating point, we locally refine
$(g_{1f},g_{2f})$, $g_{fn}$, and $\kappa_f$, together with the gate
duration. The final operating point is selected using the
phase-sensitive CZ fidelity, with leakage evaluated independently.

To retain the phase information of the entangling operation, we
reconstruct the projected quantum channel $\mathcal{E}$ on the
four-dimensional computational subspace by evolving the sixteen
operator-basis elements $|i\rangle\langle j|$, with
$i,j\in\{00,01,10,11\}$. For the ideal operation
$U_{\rm CZ}=\mathrm{diag}(1,1,1,-1)$, the leakage-aware average CZ
fidelity is
\begin{equation}
F_{\rm avg}^{\rm CZ}
=
\frac{
\sum_i {\rm Tr}\!\left[\mathcal{E}(|i\rangle\langle i|)\right]
+
\sum_{i,j}
\langle U_{\rm CZ}i|
\mathcal{E}(|i\rangle\langle j|)
|U_{\rm CZ}j\rangle
}
{d(d+1)}, \qquad d=4 ,
\label{eq:phase_sensitive_fidelity}
\end{equation}
after optimization over independent local-$Z$ rotations on the two
qubits. This metric retains computational coherences and therefore
distinguishes a CZ gate from a population-preserving operation with
incorrect phases; for example, an ideal CZ gives
$F_{\rm avg}^{\rm CZ}=1$, whereas identity evaluated against CZ gives
$F_{\rm avg}^{\rm CZ}=0.4$.

Leakage for computational-basis input $i$ is defined as
\begin{equation}
L_i=1-{\rm Tr}\!\left[
P_{\rm bare}\mathcal{E}(|i\rangle\langle i|)
\right],
\end{equation}
with
$L_{\rm avg}=\frac14\sum_iL_i$ and
$L_{\rm max}=\max_iL_i$.
For the architecture and coherence comparisons, we additionally use
the population-retention diagnostic
\[
P_{{\rm ret},i}
=
\langle i_{\rm CZ}|
\mathcal{E}(|i\rangle\langle i|)
|i_{\rm CZ}\rangle ,
\]
where $|i_{\rm CZ}\rangle=U_{\rm CZ}|i\rangle$, and report
$P_{\rm ret,avg}=\frac14\sum_iP_{{\rm ret},i}$ and
$P_{\rm ret,min}=\min_iP_{{\rm ret},i}$.
Unlike $F_{\rm avg}^{\rm CZ}$, these population-retention quantities
do not diagnose coherent phase errors.

\begin{figure}[t]
\centering
\safeincludegraphics[width=0.49\columnwidth]{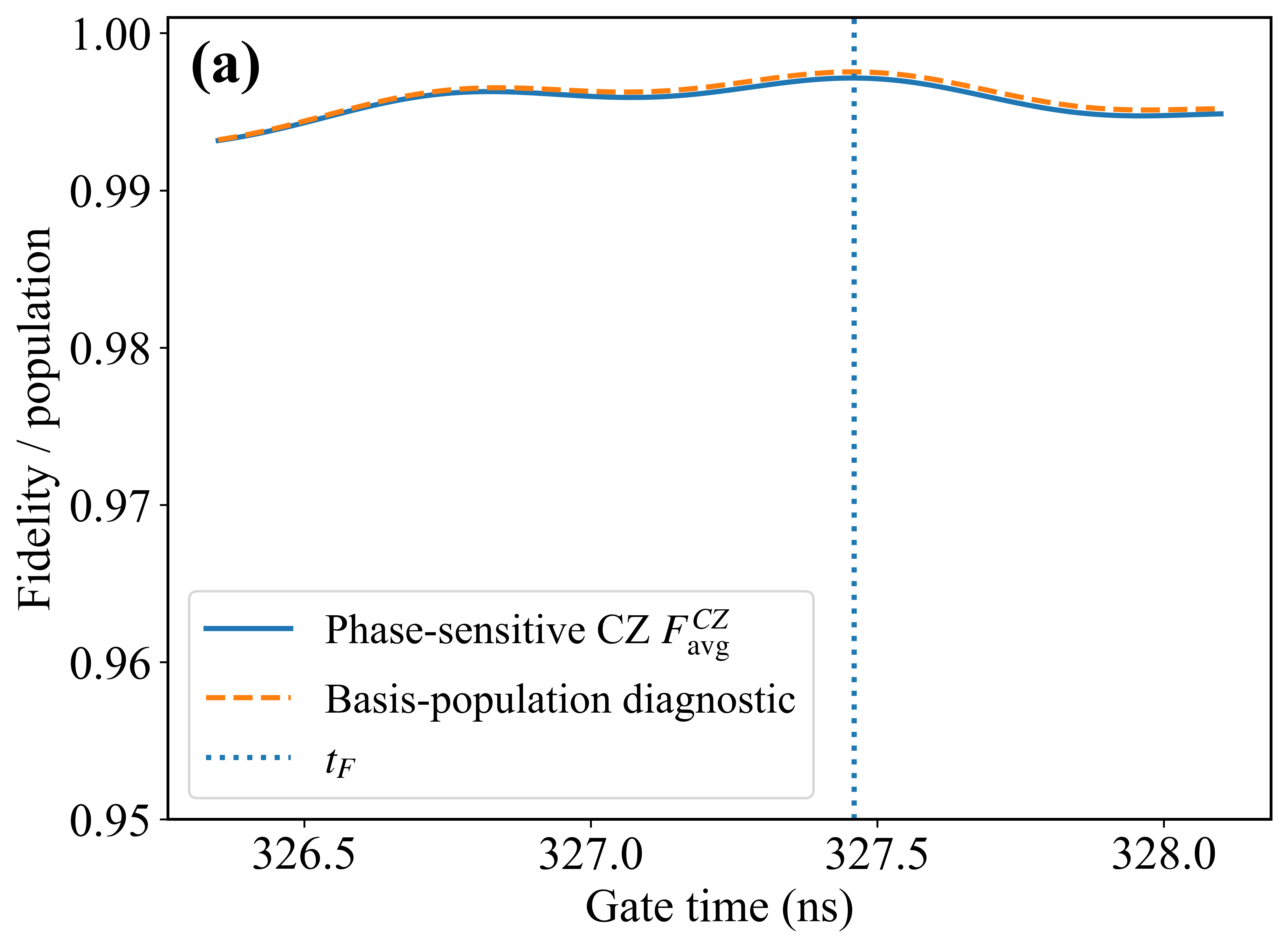}
\safeincludegraphics[width=0.49\columnwidth]{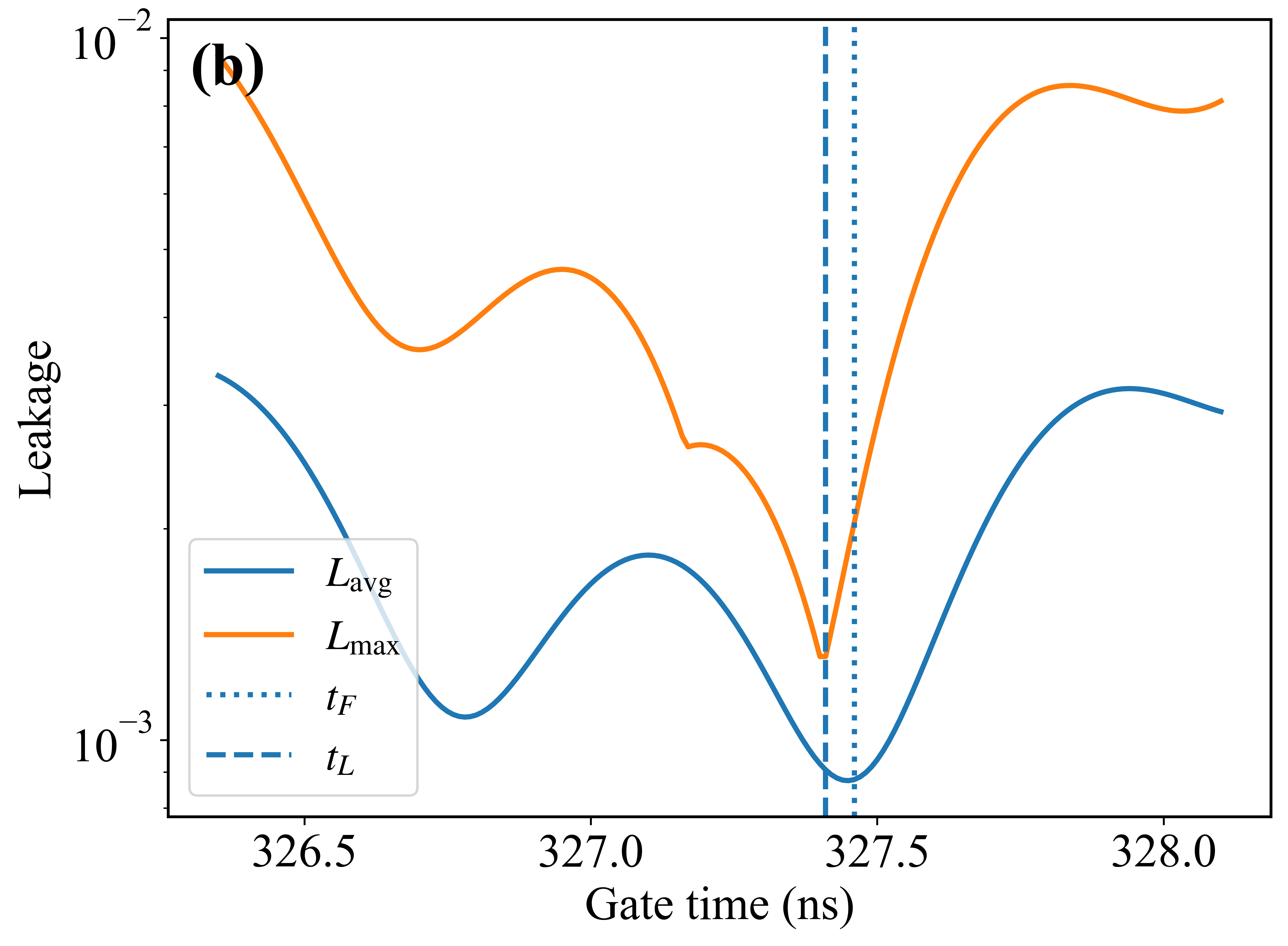}
\caption{\label{fig:local_refinement}
Phase-sensitive local optimization of the hybrid CZ gate.
(a) Projected-channel average CZ fidelity $F_{\rm avg}^{\rm CZ}$ as a
function of gate duration; the computational-state population-retention diagnostic
$P_{\rm ret,avg}$ is shown only for comparison and is not used as the CZ
fidelity metric.
(b) Average leakage $L_{\rm avg}$ and maximum computational-basis
leakage $L_{\rm max}$.  The phase-sensitive fidelity is maximized at
$t_F=327.46~\mathrm{ns}$, where
$F_{\rm avg}^{\rm CZ}=0.997144$,
$L_{\rm avg}=8.78\times10^{-4}$, and
$L_{\rm max}=2.04\times10^{-3}$.}
\end{figure}

Figure~\ref{fig:local_refinement} shows the phase-sensitive time optimization.
At the nominal optimum $t_F=327.46~\mathrm{ns}$, the conditional phase is
$-3.06696~\mathrm{rad}$, corresponding to a wrapped phase error
$|\Delta\phi_{\rm CZ}|=7.46\times10^{-2}~\mathrm{rad}$.
An independent random-pure-state check gives a sampled mean state
fidelity of $0.997126$ and a sampled minimum of $0.994941$ at this
operating point; the corresponding projected-channel fidelity and leakage
values are given in Fig.~\ref{fig:local_refinement} and Table~\ref{tab:comparison}.

\begin{figure}[t]
\centering
\safeincludegraphics[width=0.49\columnwidth]{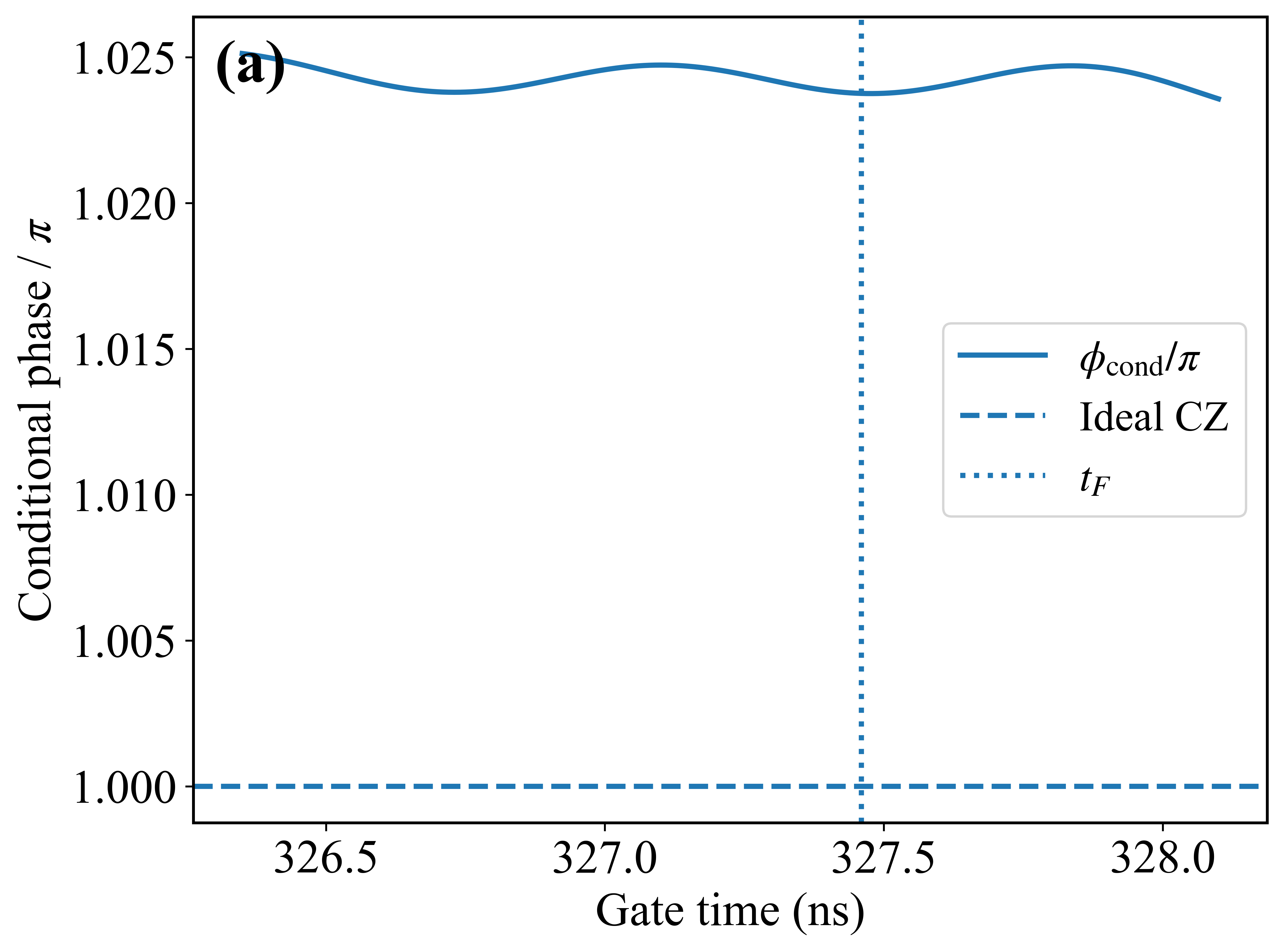}
\safeincludegraphics[width=0.49\columnwidth]{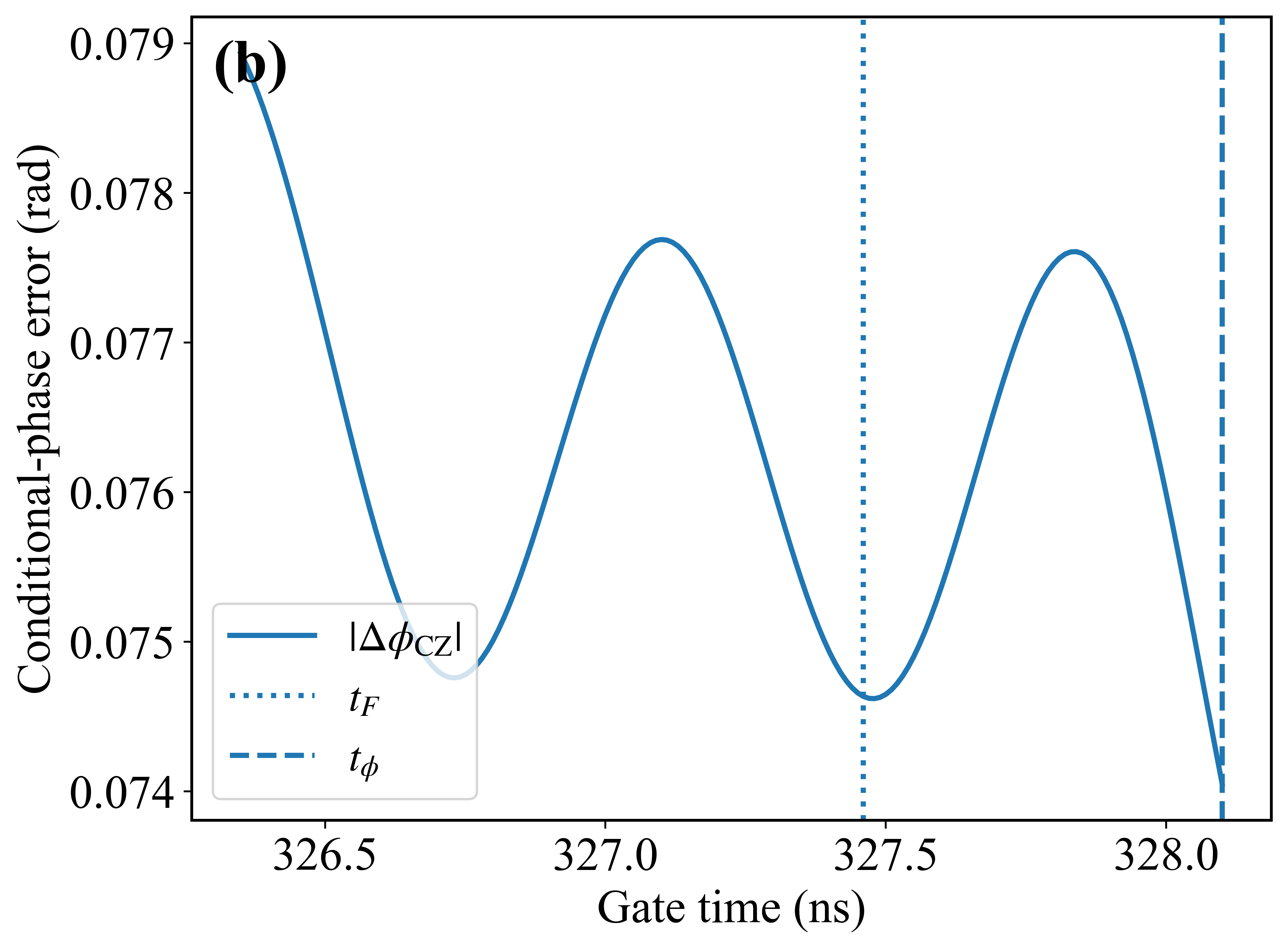}
\caption{\label{fig:phase_tradeoff}
Phase and error tradeoff near the CZ operating point.
(a) Wrapped conditional-phase error
$|\Delta\phi_{\rm CZ}|$ versus gate duration.
(b) Phase-sensitive CZ infidelity $1-F_{\rm avg}^{\rm CZ}$,
maximum computational-basis leakage $L_{\rm max}$, and normalized
conditional-phase error. 
}
\end{figure}

As shown in Fig.~\ref{fig:phase_tradeoff}, the leakage minimum occurs
nearby at $t_L=327.41~\mathrm{ns}$, where
$L_{\rm max}=1.31\times10^{-3}$ and
$F_{\rm avg}^{\rm CZ}=0.997094$. Thus, the high-fidelity and
low-leakage regions nearly coincide, although their exact extrema need
not occur at the same interaction time. The gate is therefore selected
by the complete phase-sensitive channel fidelity rather than by forcing
the conditional phase to equal $\pi$ exactly, since moving toward the
exact-phase point can increase leakage and reduce the complete gate
fidelity.

\begin{figure}[t]
\centering
\safeincludegraphics[width=0.49\columnwidth]{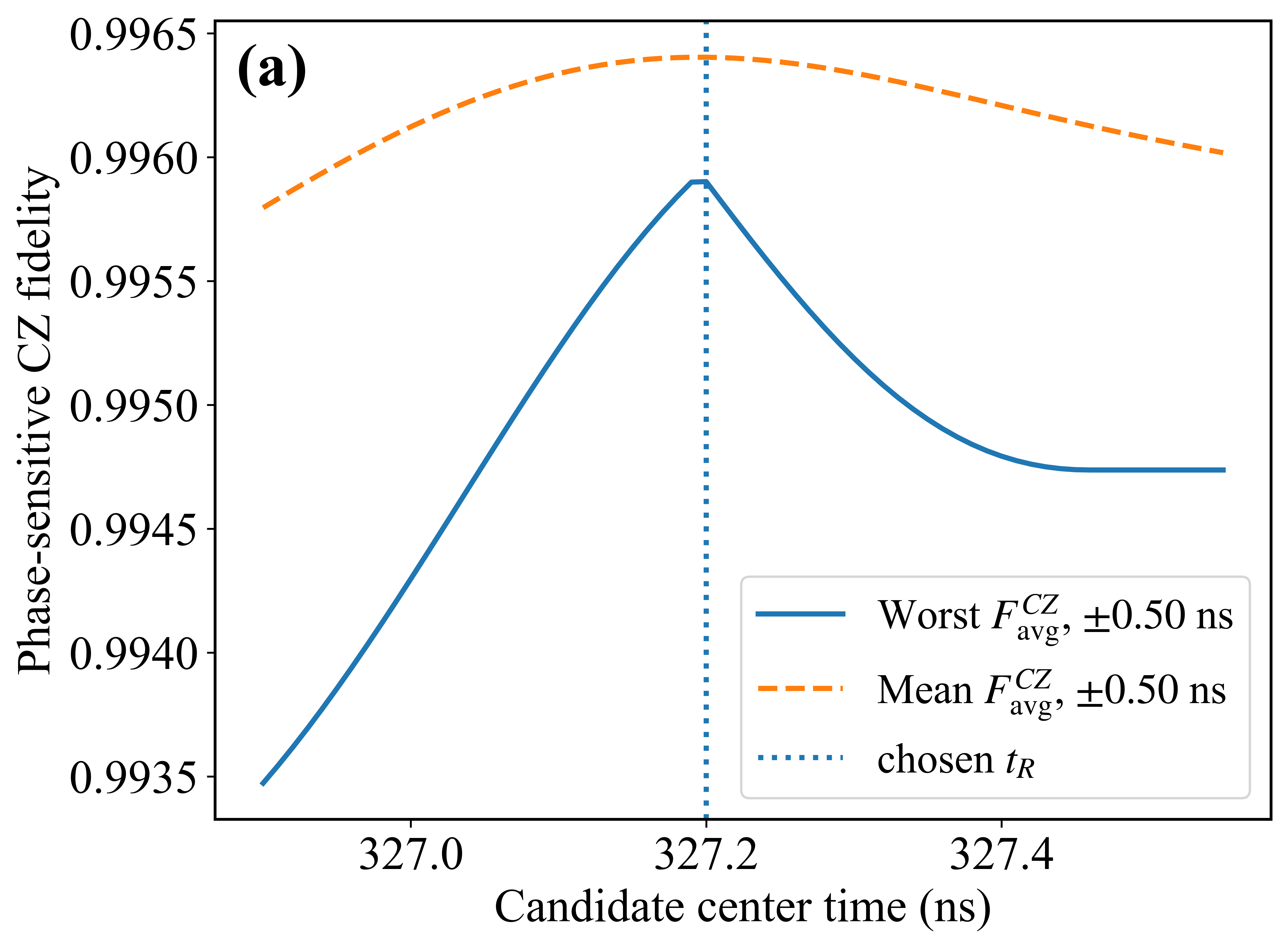}
\safeincludegraphics[width=0.49\columnwidth]{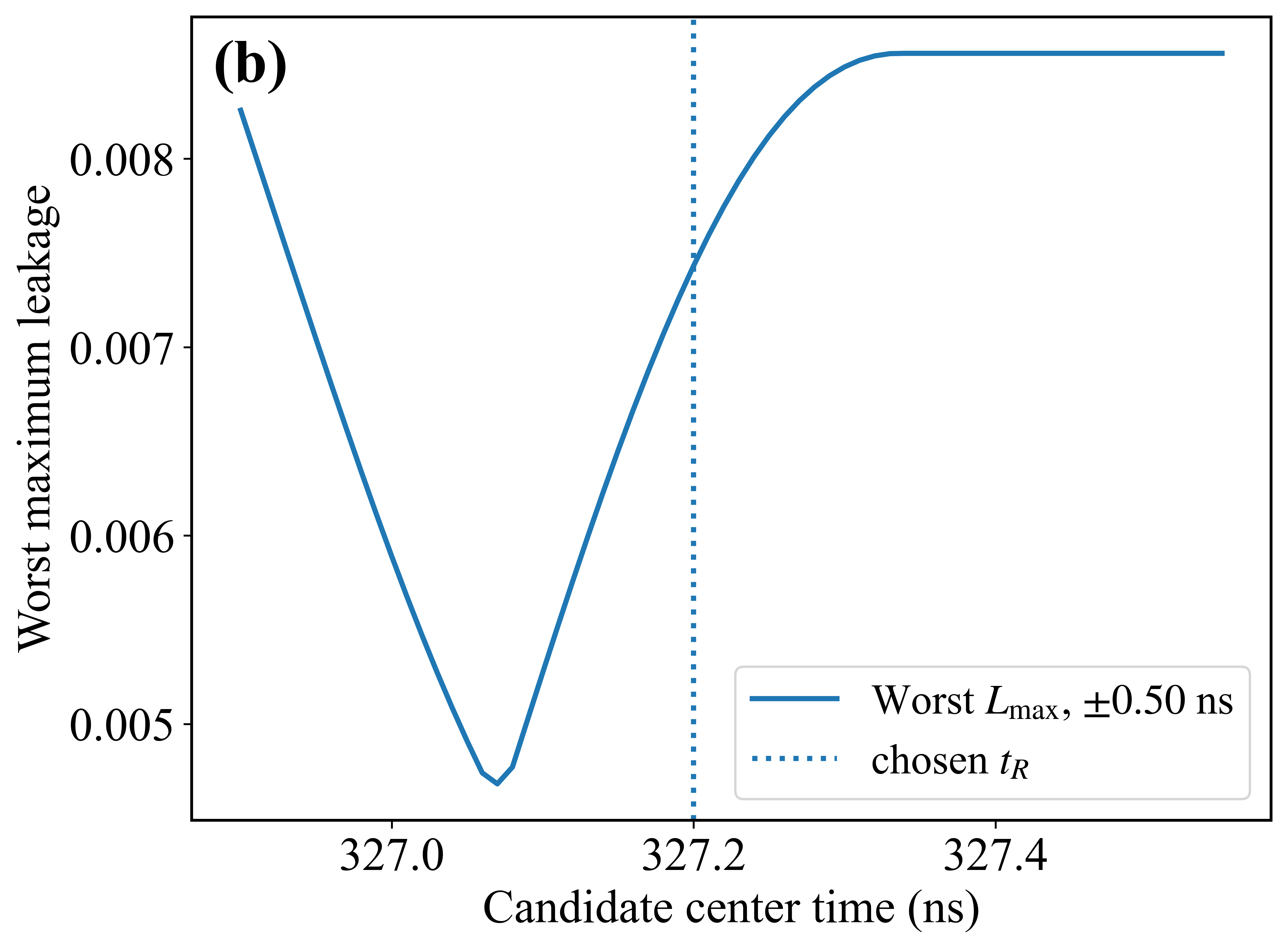}
\caption{\label{fig:timing_robustness}
Timing robustness of the phase-sensitive CZ gate.
For each candidate center $t_0$, the gate is evaluated throughout
$t_0\pm0.50~\mathrm{ns}$.
(a) Worst-case and mean $F_{\rm avg}^{\rm CZ}$ in the timing window.
(b) Worst-case maximum computational-basis leakage.
The minimax optimization selects the interior robust operating point
$t_R=327.20~\mathrm{ns}$.  Its nominal fidelity is $0.996186$;
the worst-case phase-sensitive CZ fidelity over the full timing window is $0.995901$,
and the worst-case $L_{\rm max}$ is $7.43\times10^{-3}$.}
\end{figure}

\begin{table}[t]
\caption{\label{tab:comparison}
Phase-sensitive operating points of the hybrid CZ gate.}
\begin{ruledtabular}
\begin{tabular}{lcc}
Quantity & Nominal optimum & Timing-robust point \\
Gate time & $327.46~\mathrm{ns}$ & $327.20~\mathrm{ns}$ \\
$F_{\rm avg}^{\rm CZ}$ & $0.997144$ & $0.996186$ \\
$L_{\rm avg}$ & $8.78\times10^{-4}$ & $1.66\times10^{-3}$ \\
$L_{\rm max}$ & $2.04\times10^{-3}$ & $2.63\times10^{-3}$ \\
Worst $F_{\rm avg}^{\rm CZ}$ in window & --- & $0.995901$ \\
Timing window & --- & $\pm0.50~\mathrm{ns}$ \\
\end{tabular}
\end{ruledtabular}
\end{table}

The timing scan in Fig.~\ref{fig:timing_robustness} shows that the high
phase-sensitive fidelity is not restricted to a narrow numerical revival.
The minimax point $t_R=327.20~\mathrm{ns}$ incurs a nominal fidelity
penalty of $9.58\times10^{-4}$ relative to $t_F$, while maintaining
$F_{\rm avg}^{\rm CZ}\geq0.995901$ for timing offsets up to
$\pm0.50~\mathrm{ns}$. This timing-domain test complements the
device-parameter robustness analyses below.

\subsubsection{Comparison with a Single-Transmon Coupler}

To evaluate the proposed architecture, we compare the optimized hybrid
coupler with an optimized single-transmon coupler across the coherence
sets listed in Table~\ref{tab:coherence_sets}. The qubit and coupler
relaxation and dephasing times span conservative experimental values to
a near-lossless reference, allowing coherent leakage to be distinguished
from ordinary relaxation and dephasing.

\begin{figure}[t]
\centering
\safeincludegraphics[width=0.41\textwidth]{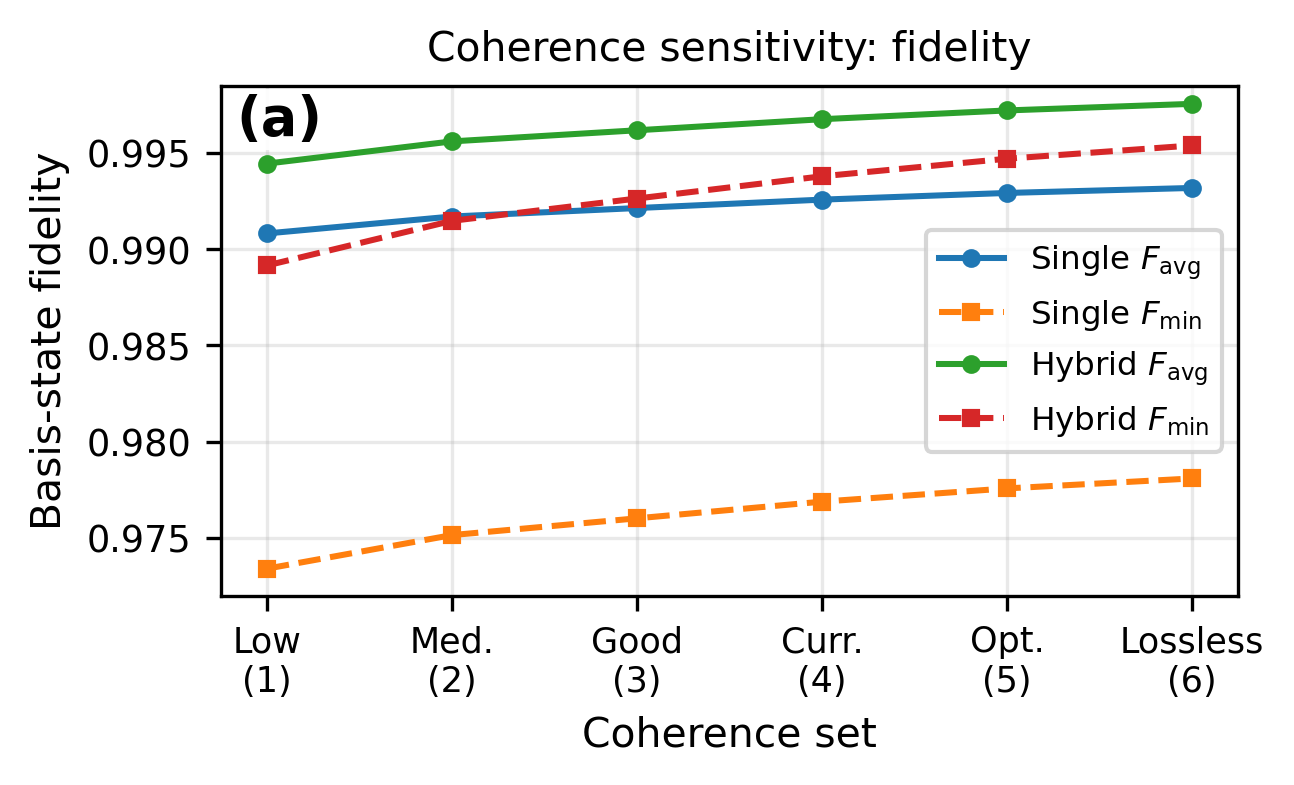}
\safeincludegraphics[width=0.41\textwidth]{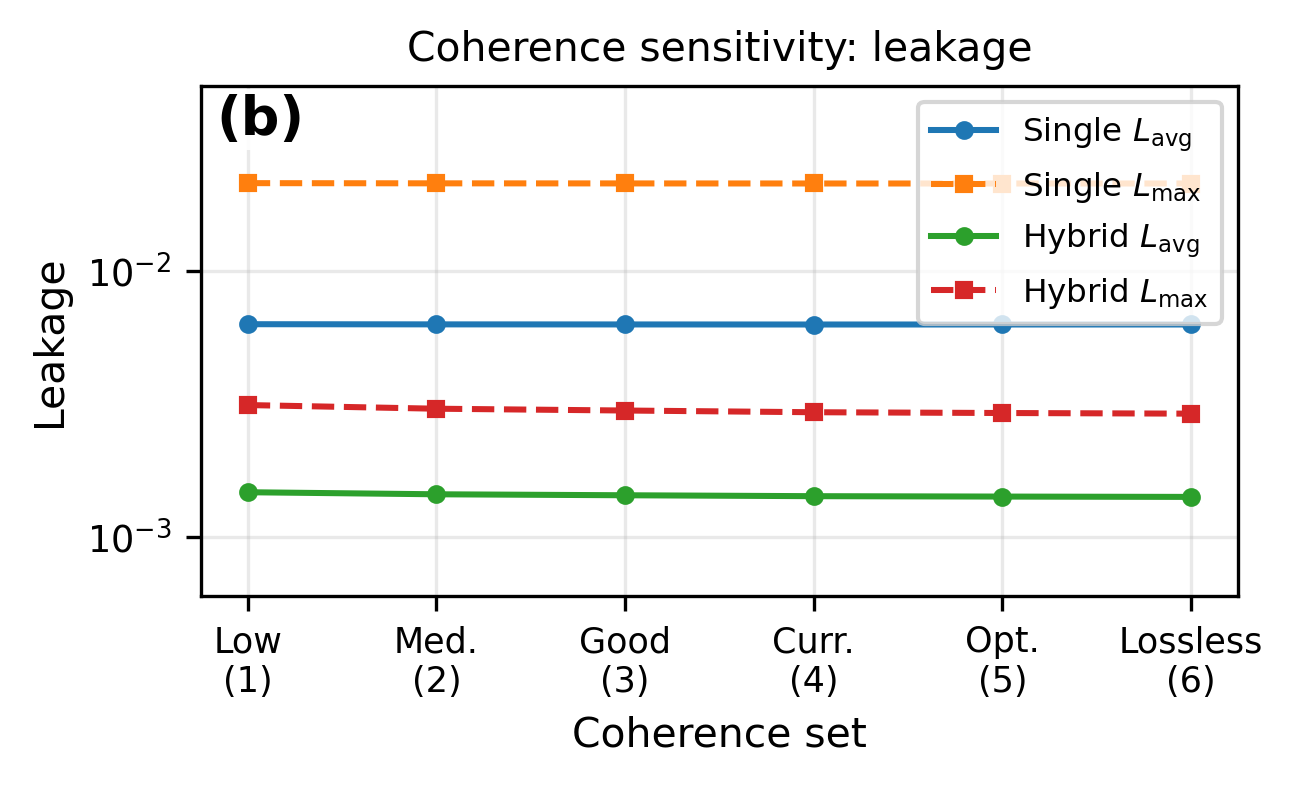}
\caption{\label{fig:coherence_compare}
Comparison between the optimized hybrid coupler and optimized
single-transmon coupler under different coherence assumptions.
(a) Average and minimum computational-state population retentions.
(b) Average and maximum leakage probabilities. The corresponding
coherence parameters are listed in Table~\ref{tab:coherence_sets}.}
\end{figure}

Figure~\ref{fig:coherence_compare} compares the two architectures over
these coherence sets; the labels Low, Med., Good, Curr., Opt., and
Lossless correspond to Sets~1--6 in Table~\ref{tab:coherence_sets}.
Panel (a) shows the average and minimum computational-state population
retentions, while panel (b) shows the corresponding leakage probabilities.
The hybrid architecture maintains higher population retention and reduces
both average and maximum leakage by approximately one order of magnitude
across the coherence range. The persistence of this reduction toward the
high-coherence limit indicates that leakage suppression is primarily
determined by the engineered filter--notch structure rather than by the
specific decoherence parameters.

\subsubsection{Role of the Filter Branch and Notch Resonator}

To distinguish the roles of the different circuit components, we compare
an optimized single-transmon coupler, a hybrid coupler with the filter
branch but without the notch resonator ($g_{fn}=0$), and the full
filter--notch hybrid coupler using the same master-equation evaluation
procedure.

\begin{figure}[t]
\centering
\safeincludegraphics[width=0.40\textwidth]{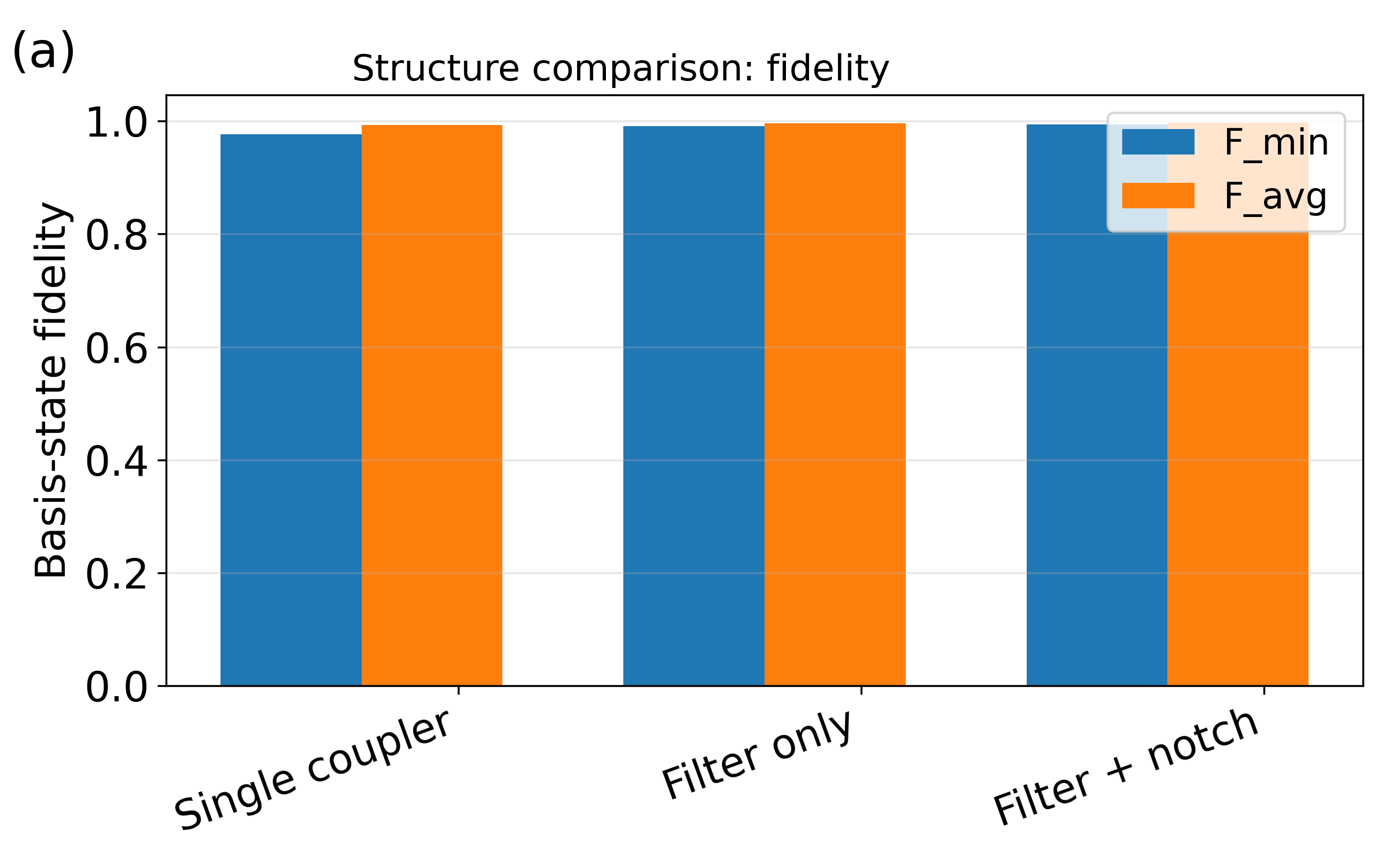}
\safeincludegraphics[width=0.41\textwidth]{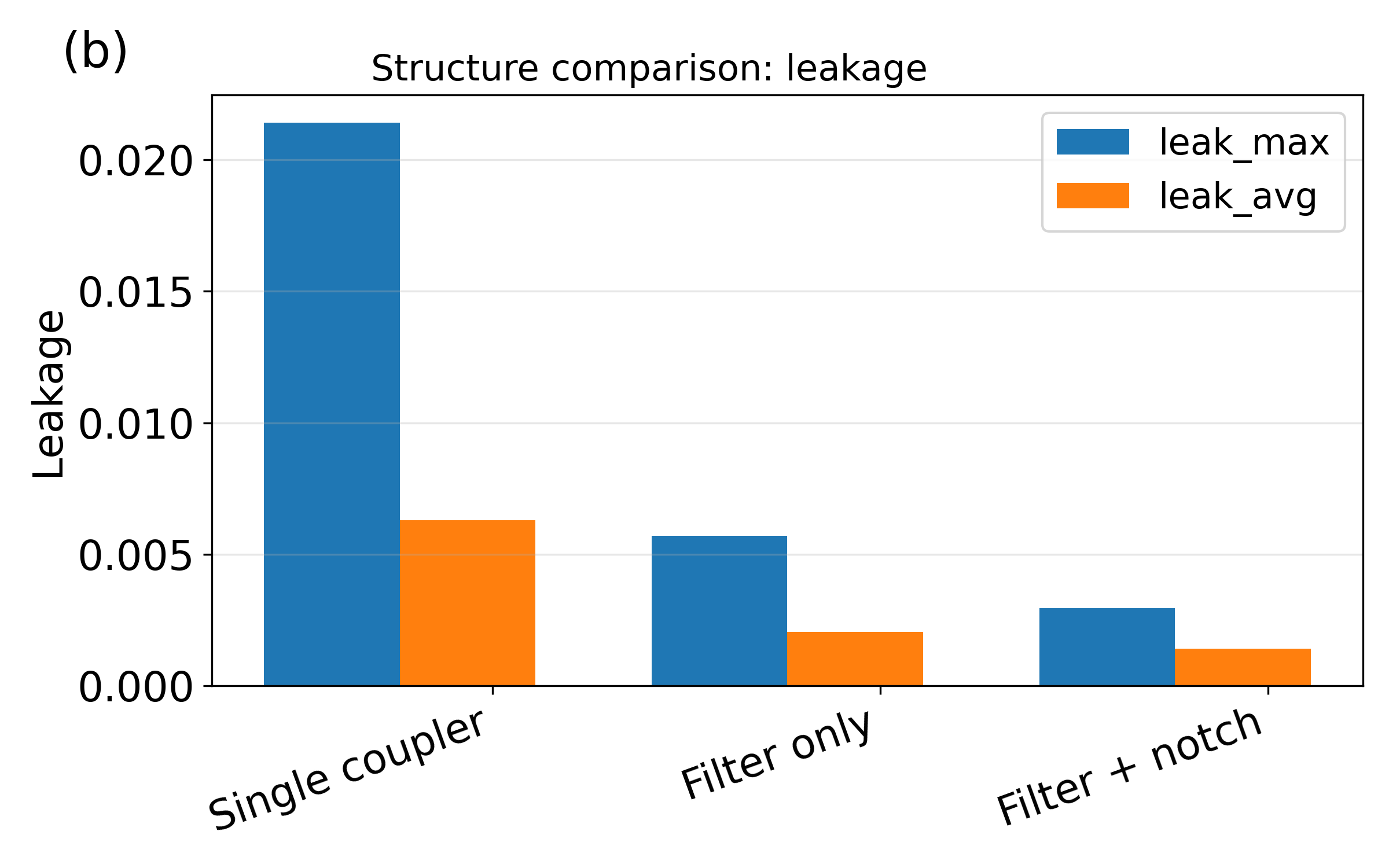}
\caption{\label{fig:structure_compare}
Structure comparison between the optimized single-transmon coupler,
the filter-only hybrid coupler with $g_{fn}=0$,
and the full filter--notch hybrid coupler.
(a) Average and minimum computational-state population retentions.
(b) Average and maximum leakage probabilities.
}
\end{figure}

\begin{table}[t]
\caption{Comparison of the three architectures evaluated using the
same simulation framework. Unlike Table~\ref{tab:comparison}, these values are obtained
from a common architecture-comparison evaluation and therefore need
not coincide exactly with the locally optimized operating point.}
\begin{tabular}{lccc}
\hline
Architecture & $P_{\rm ret,avg}$ & $P_{\rm ret,min}$ & $L_{\rm max}$ \\
\hline
Single coupler & $\sim 0.992$ & $\sim 0.975$ & $\sim 2.1\times10^{-2}$ \\
Filter only & $\sim 0.998$ & $\sim 0.991$ & $\sim 5.7\times10^{-3}$ \\
Filter + notch & $\sim 0.9974$ & $\sim 0.994$ & $\sim 3.0\times10^{-3}$ \\
\hline
\end{tabular}
\label{tab:three_arch_compare}
\end{table}

Figure~\ref{fig:structure_compare} compares the gate performance of the
three architectures, with the corresponding values summarized in
Table~\ref{tab:three_arch_compare}. Introducing the filter branch produces
the dominant reduction in leakage, decreasing $L_{\rm max}$ from
approximately $2.1\times10^{-2}$ for the optimized single-transmon
coupler to $5.7\times10^{-3}$ for the filter-only architecture. Adding
the notch resonator further reduces $L_{\rm max}$ to approximately
$3.0\times10^{-3}$. A similar trend is observed for the minimum
computational-state population retention, which improves from approximately $0.975$
for the single-transmon coupler to $0.991$ for the filter-only
architecture and $0.994$ for the full filter--notch design. These
results indicate that the filter branch provides the dominant leakage
reduction, while the notch resonator delivers an additional improvement
in both leakage suppression and minimum population retention.

\subsection{Robustness}

We evaluate the robustness of the optimized hybrid coupler against
variations in the primary filter and notch parameters.
For each parameter set, $J_{ZZ}$ is re-extracted from numerical
diagonalization, and the gate duration is determined from a local
master-equation time sweep.
The resulting parameter maps are characterized by the minimum
computational-state population retention $P_{\rm ret,min}$ and maximum
leakage probability $L_{\rm max}$. These quantities complement the
phase-sensitive timing analysis of Sec.~III.B.1.

\subsubsection{Robustness Against Filter Parameters}

Unlike the auxiliary notch resonator, the filter couples directly to
the nonlinear transmon coupler through $g_{cf}$ and indirectly to the
qubits through the weak qubit--filter couplings.
Consequently, variations in $\omega_f$ and $g_{cf}$ modify both the
coherent interaction pathways and the engineered dissipative
environment.
 We first examine the robustness with respect to the primary filter
parameters, namely the Purcell-filter frequency $\omega_f$ and the
coupler--filter coupling strength $g_{cf}$. To quantify this sensitivity, we perform a two-dimensional scan over
$(\omega_f,g_{cf})$ around the optimized operating point.
At each scan point, the dressed spectrum is recalculated,
$J_{ZZ}$ is re-extracted, and the gate duration is obtained from a
local time sweep around $t_{\rm CZ}$.
The resulting population-retention diagnostic and leakage are then
evaluated using the master-equation simulation.

\begin{figure}[t]
\centering
\safeincludegraphics[width=0.23\textwidth]{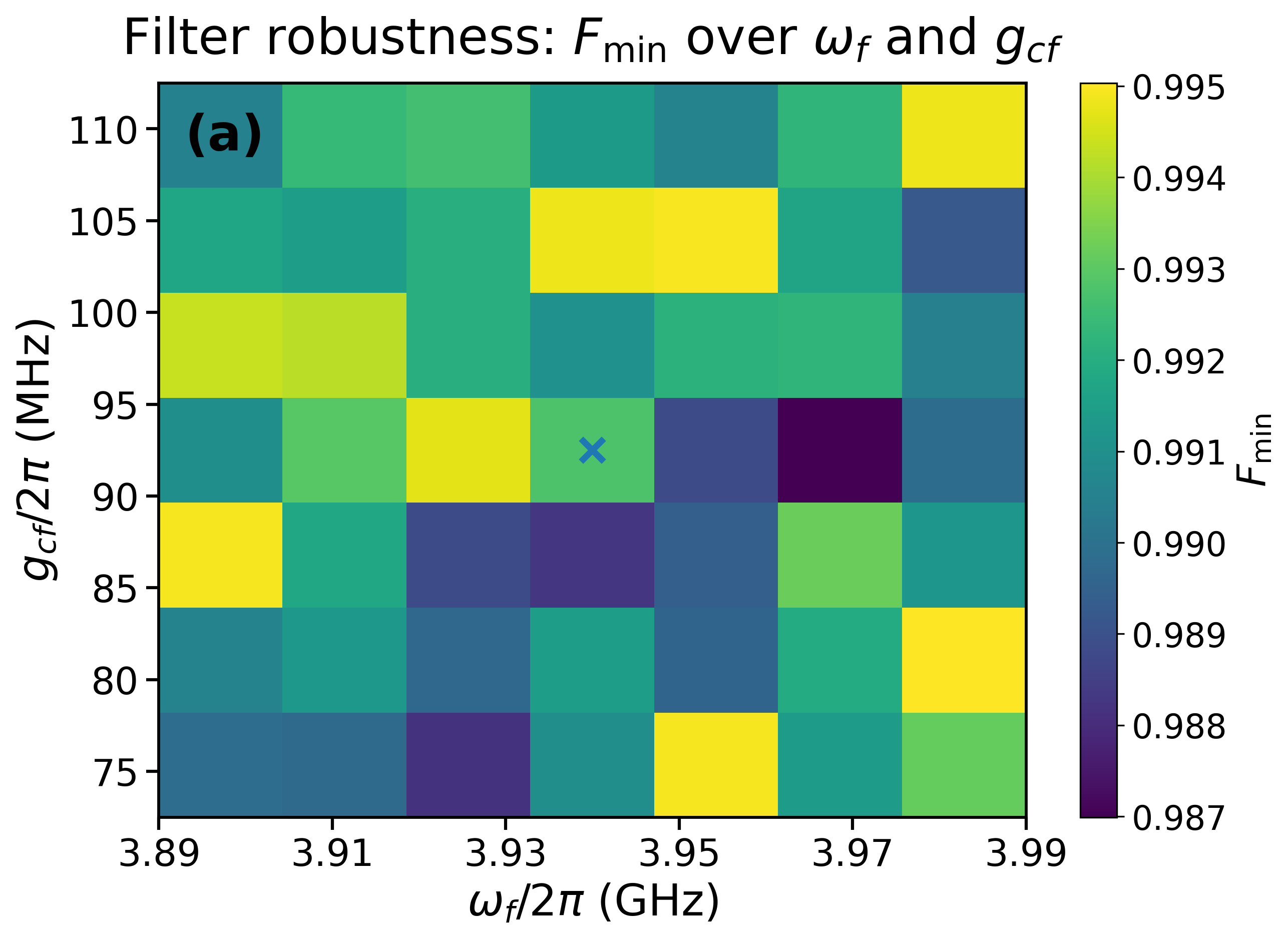}
\safeincludegraphics[width=0.23\textwidth]{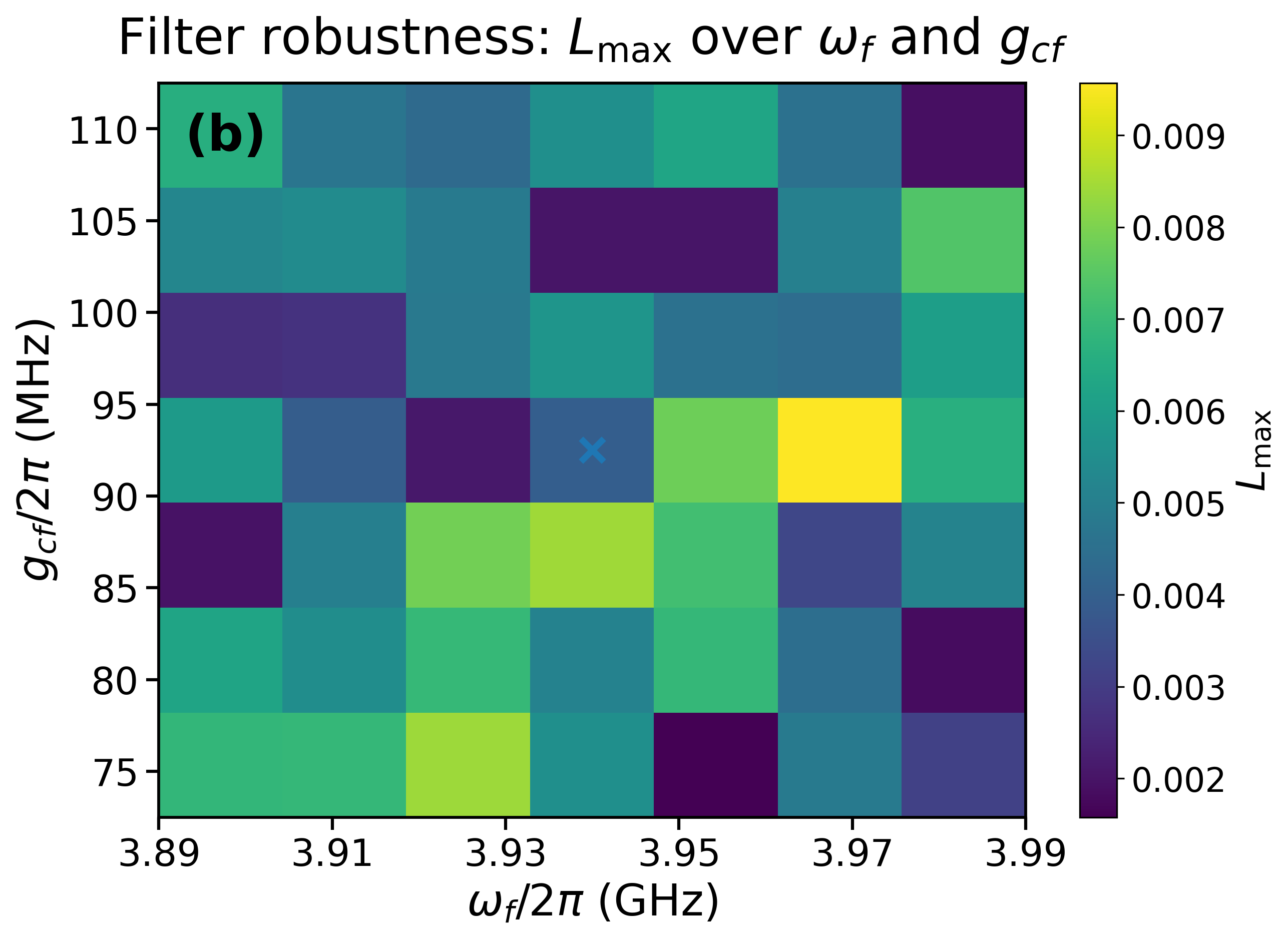}
\caption{\label{fig:wf_gcf_robustness}
Robustness of the optimized hybrid coupler against variations
of the Purcell-filter frequency $\omega_f$ and the coupler--filter coupling $g_{cf}$.
(a) Minimum computational-state population retention $P_{\rm ret,min}$.
(b) Maximum leakage probability $L_{\max}$.
The marker indicates the optimized operating point used in the main simulations.
For each scan point, $J_{ZZ}$ is re-extracted, and the gate
duration is locally refined by a discrete time sweep around
the estimate $t_{\rm CZ}\approx1/(2|J_{ZZ}|)$.
 }
\end{figure}

Figure~\ref{fig:wf_gcf_robustness}(a) shows the resulting minimum
computational-state population retention $P_{\rm ret,min}$.
Across the scanned region,
$P_{\rm ret,min}$ remains above approximately $0.987$ and reaches values as
high as $\sim0.995$ near the optimized operating point.
Figure~\ref{fig:wf_gcf_robustness}(b) shows the corresponding maximum
leakage probability $L_{\max}$, which remains below
$\sim10^{-2}$ throughout the scanned parameter range.
Compared with the notch-parameter scans, the
$(\omega_f,g_{cf})$ landscape is less smooth because these parameters
directly modify both the coherent interactions and the engineered
dissipative environment.
Nevertheless, a broad region of simultaneously high
$P_{\rm ret,min}$ and low
$L_{\max}$ demonstrates that the optimized gate performance does not
depend on precise tuning of the filter parameters.

\subsubsection{Robustness Against Notch Parameters}

We next examine variations in the notch frequency $\omega_n$ and the
filter--notch coupling $g_{fn}$.  Because the notch resonator couples only to the Purcell filter, these
parameters primarily reshape the frequency response of the coupled
filter--notch subsystem while only weakly affecting the coherent
interactions. At each scan point, $J_{ZZ}$ is
re-extracted and the gate duration is locally optimized before
evaluating the performance metrics.

\begin{figure*}[t]
\centering
\safeincludegraphics[width=0.31\textwidth]{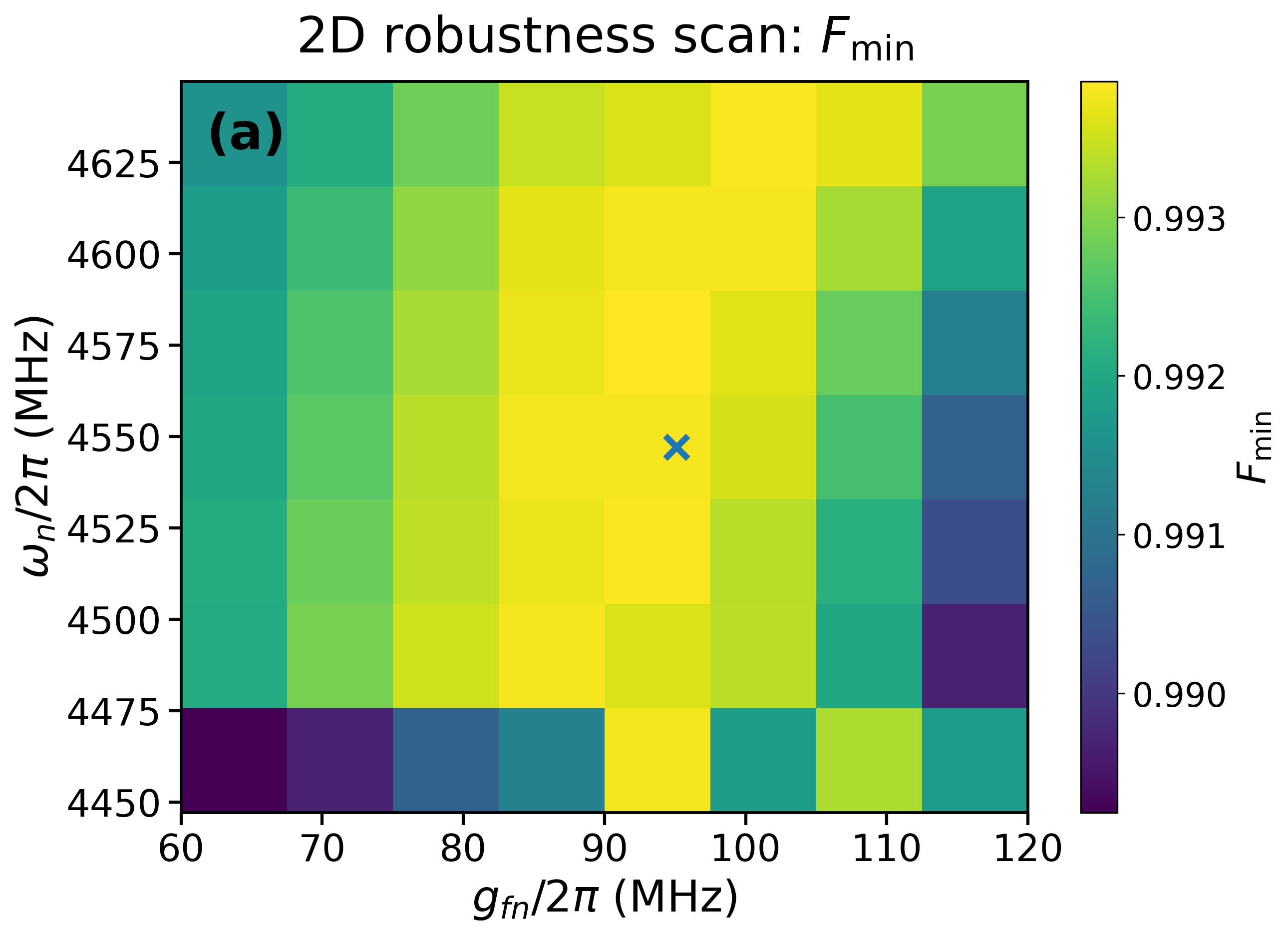}
\safeincludegraphics[width=0.31\textwidth]{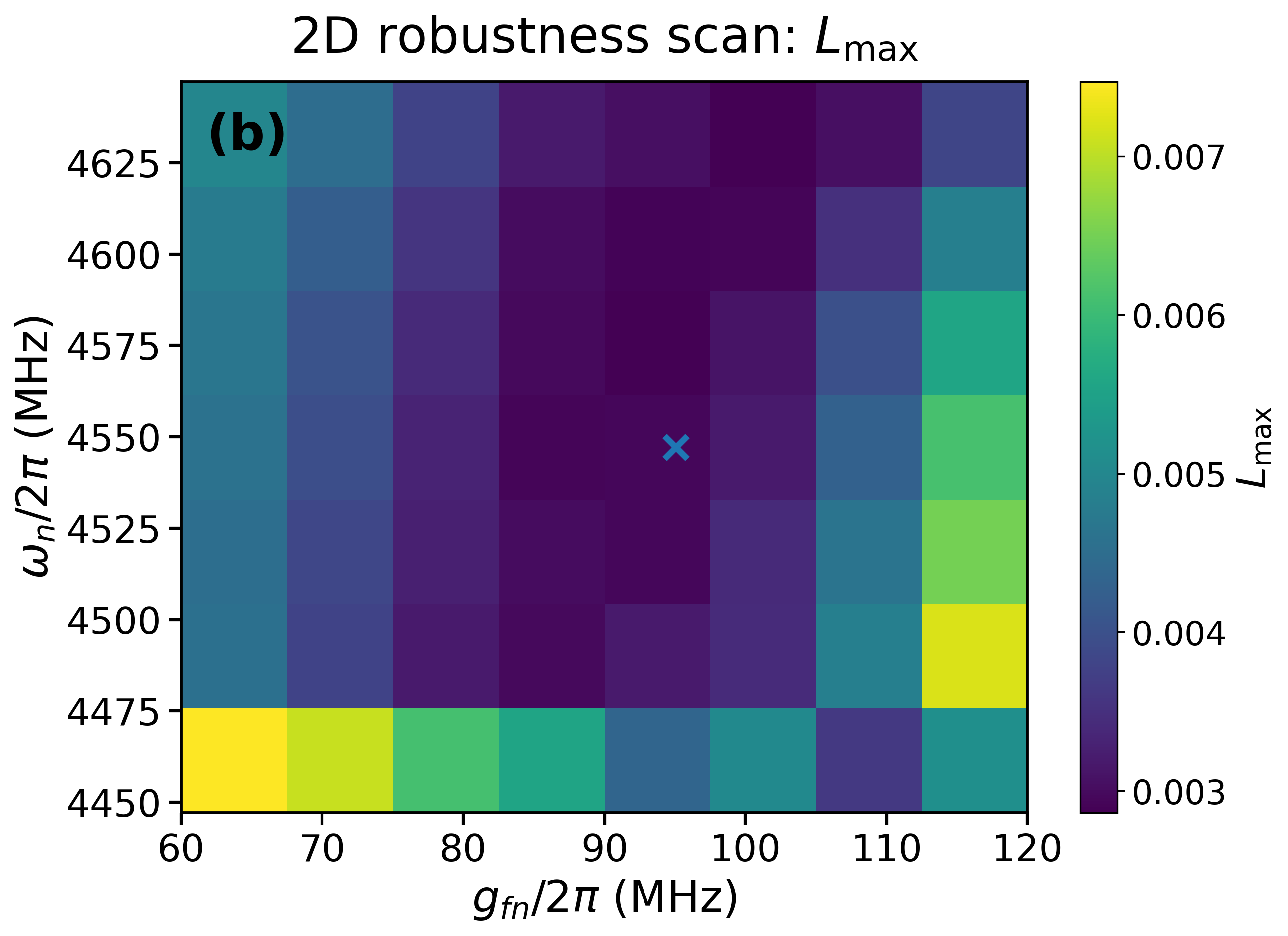}
\safeincludegraphics[width=0.31\textwidth]{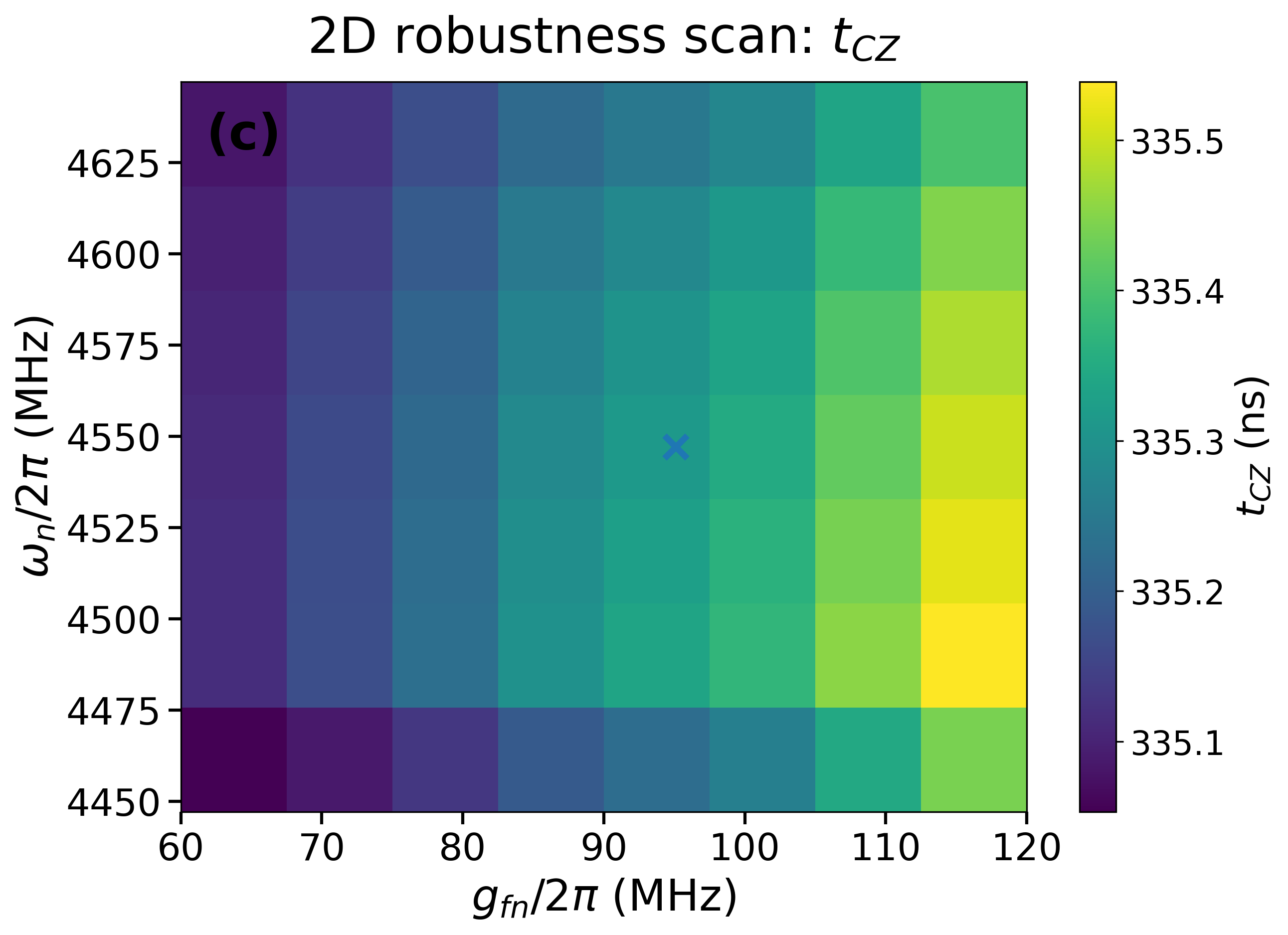}
\caption{\label{fig:gfn_wn_map}
Robustness of the full filter--notch hybrid coupler against
variations of the notch frequency $\omega_n$ and the filter--notch
coupling $g_{fn}$.
(a) Minimum computational-state population retention $P_{\rm ret,min}$.
(b) Maximum leakage probability $L_{\max}$.
(c) Optimized CZ-gate duration $t_{\rm CZ}$ obtained after
re-extracting $J_{ZZ}$ and locally optimizing the interaction
time at each scan point.
The marker indicates the optimized operating point used in the
main simulations.
The weak variation of $t_{\rm CZ}$ across the scanned region
indicates that the dominant conditional interaction remains
primarily controlled by the nonlinear transmon coupler, while
the notch subsystem mainly modifies leakage dynamics and
engineered dissipation pathways.
}
\end{figure*}

Figure~\ref{fig:gfn_wn_map}(a) shows a broad high-retention region
around $g_{fn}/2\pi\simeq85$--$105~{\rm MHz}$ and
$\omega_n/2\pi\simeq4.520$--$4.620~{\rm GHz}$, with the optimized
operating point near its center. The corresponding low-leakage region
in Fig.~\ref{fig:gfn_wn_map}(b) largely overlaps this region,
demonstrating that the favorable operating regime is robust against
moderate variations of the notch parameters.

As shown in Fig.~\ref{fig:gfn_wn_map}(c), $t_{\rm CZ}$ varies by less
than approximately $0.5~{\rm ns}$ across the scanned region. Thus,
variations in $\omega_n$ and $g_{fn}$ have only a weak influence on
the conditional interaction while substantially modifying the
leakage behavior through the engineered dissipative environment.

\subsubsection{Robustness Against Qubit--Filter Coupling Variations}

A practical implementation also requires robustness against
fabrication-induced variations in the weak direct qubit--filter
couplings. To evaluate this sensitivity, we perform a fine asymmetric
scan of $g_{1f}$ and $g_{2f}$ around the optimized operating point,
while keeping all other coherent parameters fixed.

\begin{figure}[t]
\centering
\safeincludegraphics[width=0.494\columnwidth]{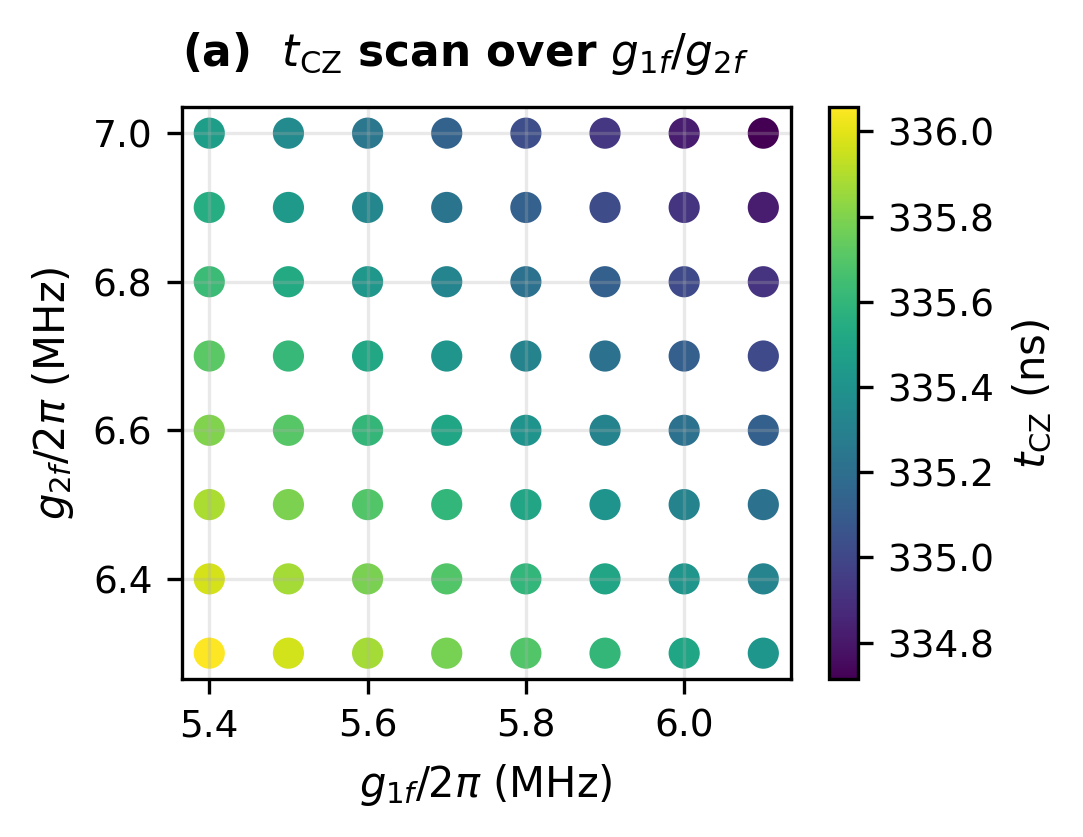}
\safeincludegraphics[width=0.494\columnwidth]{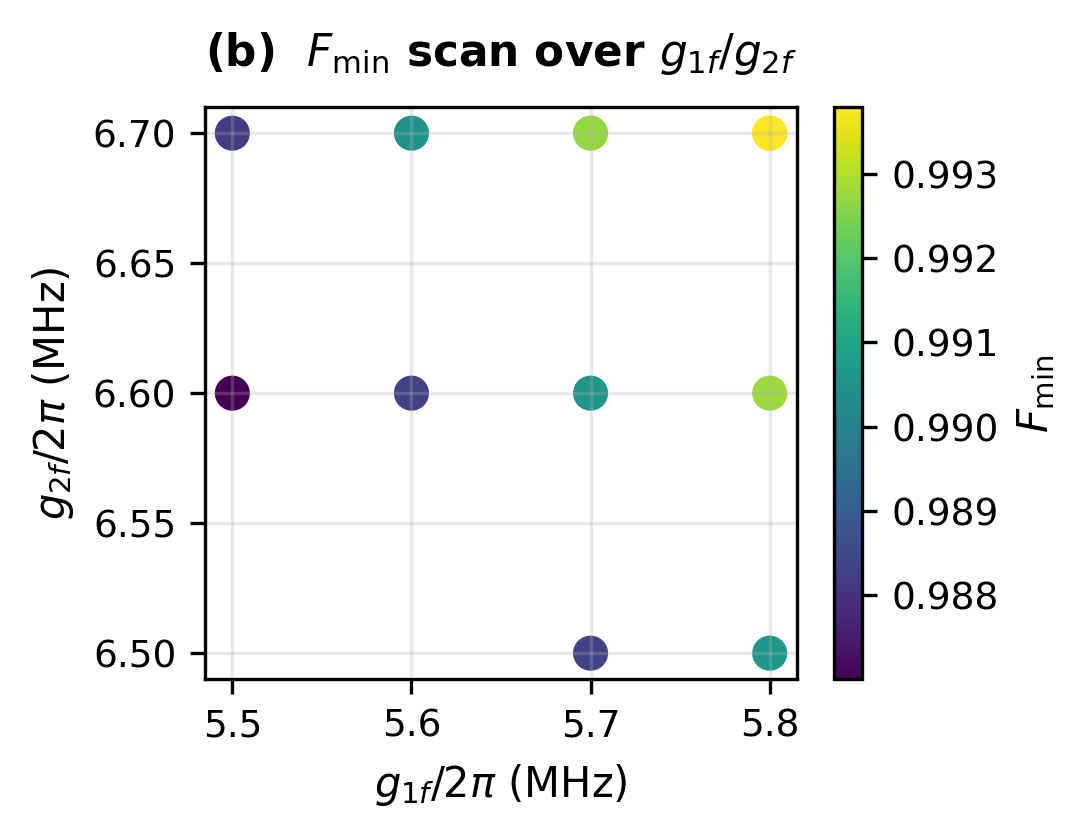}
\caption{\label{fig:g1fg2f_scan}
Robustness of the optimized hybrid coupler against variations of the qubit-filter couplings.
Panel (a) shows the dense static $J_{ZZ}$-based duration
scan, while panel (b) shows the smaller set of points evaluated
with the full master-equation simulation.
}
\end{figure}

Figure~\ref{fig:g1fg2f_scan}(a) shows the estimated static CZ-gate
duration across the scanned $g_{1f}$--$g_{2f}$ parameter space.
At each scan point, $J_{ZZ}$ is re-extracted from numerical
diagonalization, and the corresponding gate duration is estimated as
$t_{\rm CZ}\approx1/(2|J_{ZZ}|)$.
Since $g_{1f}$ and $g_{2f}$ modify the filter-mediated exchange pathway,
they slightly alter the effective qubit--qubit interaction and hence the
estimated gate duration.
Across the scanned region,
$t_{\rm CZ}$ varies only from approximately
334.8 to 336.0 ns, indicating a relatively flat parameter landscape.
This static estimate is slightly longer than the
master-equation optimized value of 327.3 ns reported in
Sec.~III because the latter is obtained by directly maximizing the phase-sensitive gate
fidelity under the full dissipative dynamics rather than from the
approximate relation based solely on $J_{ZZ}$.

Figure~\ref{fig:g1fg2f_scan}(b) shows the corresponding minimum
computational-state population retention obtained from the master-equation
simulations.  A broad high-retention region is observed around the optimized
operating point. Variations of several hundred kilohertz in the qubit--filter
couplings produce only modest changes in $P_{\rm ret,min}$, demonstrating
substantial tolerance to fabrication uncertainties.
This broad operating region indicates that the optimized design is not
highly sensitive to small variations in the qubit--filter couplings.

\section{Discussion}

The proposed architecture extends multichannel-coupler concepts by
combining coherent-interaction engineering with leakage-selective
dissipation. Related coupler-assisted leakage-reduction experiments
have transferred unwanted qubit and coupler excitations into readout
resonators, from which they dissipate through a broadband
Purcell-filter environment \cite{Yang2024LeakageReduction}. In
contrast, the present architecture incorporates a directly coupled
filter--notch subsystem into the CZ-gate circuit and exploits its
structured frequency response during gate operation. The Purcell-filter
branch provides the primary reduction in residual leakage through
modified coherent interaction pathways, state hybridization, and
coupling to the engineered dissipative environment, while the notch
resonator further shapes the environmental response to protect
computational transitions and extend the low-leakage operating region.

The additional hardware consists of passive linear resonators and their
coupling network, with resonance frequencies determined primarily by
lithographic geometry rather than Josephson-junction parameters. This
provides additional design flexibility without introducing another
nonlinear circuit element. The present simulations nevertheless assume
a fixed gate protocol with optimized device parameters and interaction
duration rather than leakage-aware optimal-control pulses. Moreover,
although shortening the lifetime of residual leakage may reduce its
persistence into subsequent operations, establishing such a multicycle
benefit requires repeated-gate or quantum-error-correction-cycle
simulations beyond the present single-gate analysis. The reported
results should therefore be interpreted as the performance of the
proposed hardware architecture under the control assumptions considered
here rather than as its ultimate performance limit.

\section{Conclusion}

We have proposed a Purcell-engineered filter--notch hybrid coupler for
superconducting CZ gates that combines coherent interaction engineering
with leakage-selective dissipation. Dressed-state analysis and
master-equation simulations show that the engineered environment
preferentially couples to leakage-related dressed states while
maintaining protection of the computational manifold.

Using the phase-sensitive projected-channel CZ fidelity, the nominal
operating point occurs at $t_{\rm CZ}=327.46~\mathrm{ns}$ with
$F_{\rm avg}^{\rm CZ}=0.997144$,
$L_{\rm avg}=8.78\times10^{-4}$, and
$L_{\rm max}=2.04\times10^{-3}$. A timing-robust operating point at
$327.20~\mathrm{ns}$ maintains
$F_{\rm avg}^{\rm CZ}\geq0.995901$ over a
$\pm0.50~\mathrm{ns}$ timing window. Together with the device-parameter
robustness results, these calculations demonstrate a low-leakage,
high-fidelity operating region that is not restricted to an isolated
interaction time or finely tuned circuit parameters. Engineered
dissipation therefore provides a complementary design degree of freedom
for robust superconducting two-qubit gates.

\section{ACKNOWLEDGMENTS}

We thank  Rui Li for   valuable suggestions.
Y.J.Z. is supported by National Natural Science Foundation of China (Grant No. 62474012).
 X.-W.X. is supported by the National Natural Science Foundation of China (Grant Nos.12064010 and 12247105).

\end{document}